\documentclass{jfm}
\usepackage{graphicx,natbib, amsmath}
\usepackage{epstopdf, epsfig}

\newcommand{\nn}{\noindent}

\newcommand{\barint}{~-\hspace{-.35cm}\int}

\shorttitle{Optimal convection cooling flows in general geometries}
\shortauthor{S. Alben}

\title{Optimal convection cooling flows in general geometries}

\author{S. Alben
  \corresp{\email{alben@umich.edu}}}

\affiliation{Department of Mathematics, University of Michigan,
Ann Arbor, MI 48109, USA}

\begin{document}

\maketitle

\begin{abstract}
We generalize a recent method for computing optimal 2D convection cooling flows
in a horizontal layer to a wide range of geometries, including those relevant
for technological applications. We write the problem in a conformal pair of coordinates 
which are the pure conduction temperature and its harmonic conjugate. We
find optimal flows for cooling a cylinder in an annular domain, a hot plate embedded
in a cold surface, and a channel with hot interior and cold exterior. With a constraint of
fixed kinetic energy, the optimal flows are all essentially the same in the 
conformal coordinates. In the physical coordinates, they consist of vortices
ranging in size from the length of the hot surface to a small cutoff length at the interface
of the hot and cold surfaces. With the constraint of fixed enstrophy (or fixed rate of
viscous dissipation), a geometry-dependent metric factor appears in the equations. 
The conformal coordinates are useful here because they map the problems to a rectangular 
domain, facilitating numerical solutions. With a small enstrophy budget,
the optimal flows are dominated by vortices which have the same size as the flow domain.
\end{abstract}

\section{Introduction}

Heat transfer plays a fundamental role in many problems of scientific,
environmental, and technological importance (\cite{raschke1960heat,rohsenow1985handbook,ozisik2000inverse,otero2004high,doering2006bounds,lienhard2013heat}). Heat transfer by the natural
and forced convection of fluids leads to many important
fluid dynamics problems across science and engineering (\cite{bird2007transport,bejan2013convection,lienhard2013heat}). For example, several
studies have proposed methods for increasing heat transfer efficiency
to alleviate constraints on computer processor speeds due to internal heating (\cite{nakayama1986thermal,zerby2002final,mcglen2004integrated,ahlers2011aircraft}).
One way to improve heat transfer is to change the spatial and temporal configurations of heat sources and sinks
(\cite{campbell1997optimal,da2004optimal,gopinath2005integrated}). Another, which we
pursue in this work, is to optimize a convecting fluid flow, subject to suitable constraints, 
for a given configuration of of heat sources and sinks. Similar flow problems have been approached in a variety
of ways depending on what is optimized and the assumptions about the underlying fluid flow
(\cite{karniadakis1988minimum,mohammadi2001applied,zimparov2006thermodynamic,chen2013entransy}). A related problem is
the optimal flow for the mixing of a passive scalar in a fluid (\cite{chien1986laminar,caulfield2001maximal,tang2009prediction,foures2014optimal,camassa2016optimal}).

The most standard geometries for convective heat transfer can be classified 
based on whether the convecting flows are
internal or external (\cite{rohsenow1998handbook,bird2007transport}). External flows
are used to transfer heat from the external surfaces of a 
heated object (e.g. flat plate (\cite{lienhard2013heat}), 
cylinder (\cite{karniadakis1988numerical}), or sphere (\cite{kotouvc2008loss})). 
Internal flows are typically used to transfer heat
from the internal surfaces of a heated pipe or duct (\cite{rohsenow1998handbook}).
The simplest internal flows 
are approximately unidirectional, with the flow profile depending strongly 
on the duct geometry and
whether the flow is laminar or turbulent, and developing or fully-developed
(\cite{lienhard2013heat}).

One set of recent work has studied heat transfer enhancement by modifying channel flows from a quasi-unidirectional flow. 
Obstacles such as rigid bluff bodies or oscillating plates or flags (active or
passive) are inserted into the flow, and vorticity emanates
from their separating boundary layers (\cite{fiebig1991heat,sharma2004heat,accikalin2007characterization,gerty2008fluidic,hidalgo2010heat,shoele2014computational,jha2015small}). The vortices
enhance the mixing of the advected temperature field and disrupt the
thermal and viscous boundary layers close to the heated surfaces (\cite{biswas1996numerical}). The
question is whether it is better to create 
vortical structures in a unidirectional background flow or simply
increase the speed of the unidirectional flow, if both options have
the same energetic cost.

In this work we consider the optimal flows
for convection cooling of heated surfaces with a range of geometries. We focus on steady 2D flows, which are expected to be the first step towards applying
the methods to unsteady 3D flows. We adopt the framework of \cite{hassanzadeh2014wall} which was motivated by the
problem of optimal heat transport in
Rayleigh-B{\'e}nard convection. Optimal flow solutions for Rayleigh-B{\'e}nard convection were computed by \cite{waleffe2015heat} and \cite{sondak2015optimal} and in a truncated model
(the Lorenz equations) by \cite{souza2015maximal,souza2015transport}. 
We adapt the framework of \cite{hassanzadeh2014wall} to a more general class of geometries including those relevant to 
convection cooling in technological applications.  
By using a convenient choice of coordinates, we show that the optimal
flows in a wide range of geometries are simply those found by 
\cite{hassanzadeh2014wall} mapped to the new coordinates, when a fixed
kinetic energy budget is imposed. With a fixed enstrophy budget, a geometry dependent
term enters the equations, but the new coordinates facilitate numerical solutions and
the qualitative understanding of the flow features.

%

\section{General framework and application to exterior cooling}\label{sec:model}
We consider 2D regions for which the boundaries are solid walls with 
fixed temperature ($T = 0$ or 1) or else insulated ($\partial_n T = 0$, where
$n$ is the coordinate normal to the boundary, increasing into the fluid domain). 
We attempt to find the 2D steady incompressible
flow of a given total kinetic energy which maximizes the heat transferred out of the hot boundaries (those with $T = 1$), which is also the heat transferred into
the cold boundaries (those with $T = 0$).
\cite{hassanzadeh2014wall} solved this problem when the surfaces are horizontal parallel lines extending to infinity. We extend their results to different geometries using
a change of coordinates.

\begin{figure}
  \centerline{\includegraphics[width=12cm]
  {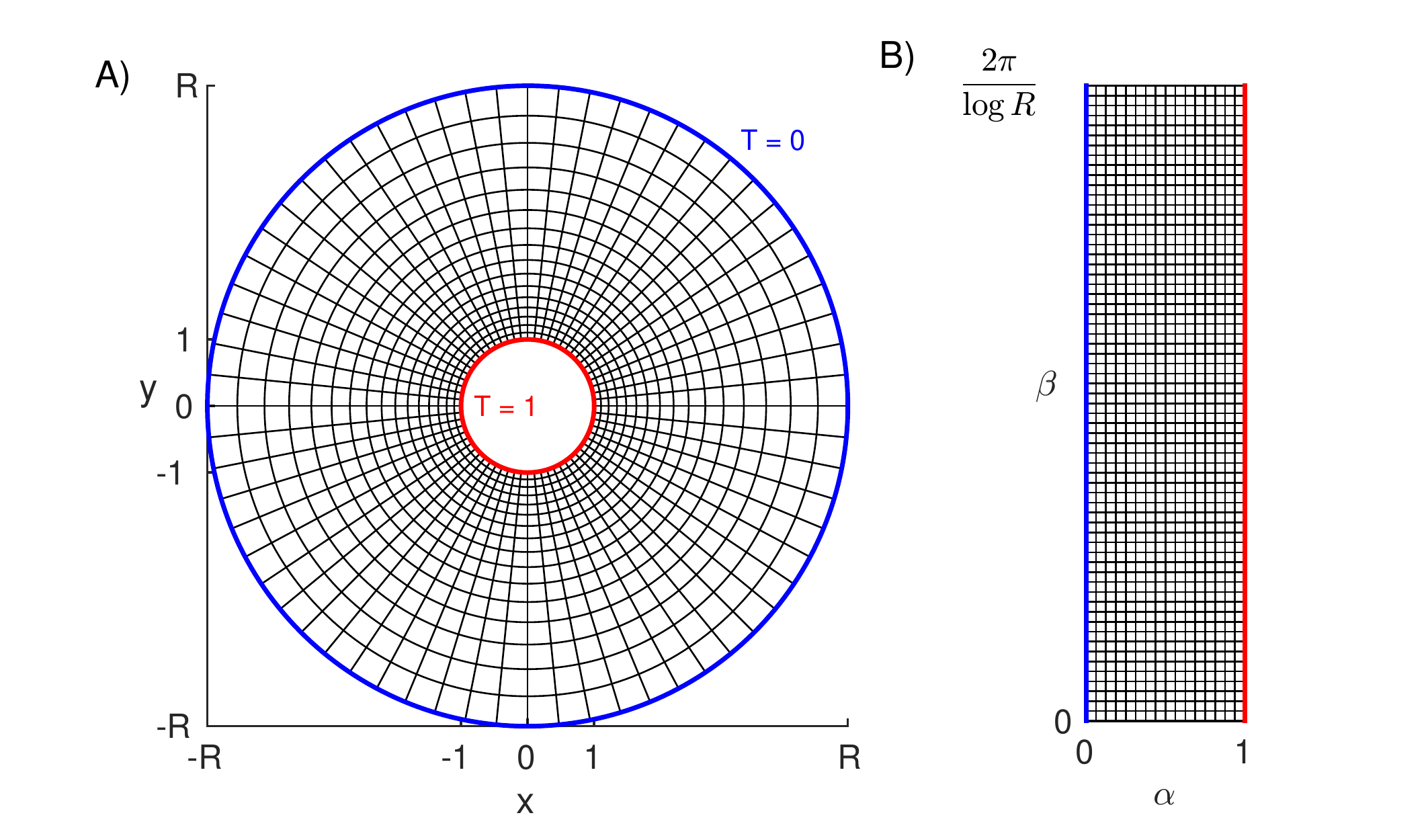}}
  \caption{A) Annular domain between the inner (hot) boundary at radius 1 and the 
outer (cold) boundary
at radius $R$. The curvilinear grid plots equicontours of the conformal coordinates  
$\alpha = 1 - \log{r}/\log{R}$ and $\beta = -\theta/\log{R}$. B) The flow domain
in ($\alpha, \beta$) coordinates.
  }
\label{fig:Annulus}
\end{figure}

We illustrate the general approach using a simple case in which the
boundaries are concentric circles (see figure \ref{fig:Annulus}).
The inner circle (radius 1) is the hot surface ($T$ = 1), and 
the outer circle (radius $R > 1$) is the cold surface ($T$ = 0), 
and can be taken to represent a cold reservoir to which heat is transferred. We solve the problem for all $R > 1$, but the typical application would be
the cooling of the exterior surface of an isolated heated body (\cite{karniadakis1988numerical}), in which case the reservoir is far away ($R \gg 1$).   

The flow $\mathbf{u}(x,y) = (u(x,y), v(x,y))$ satisfies incompressibility
($\nabla \cdot \mathbf{u} = 0$) so we write it in terms of the stream function $\psi(x,y)$ as
$\mathbf{u} = -\nabla^\perp \psi = (\partial_y \psi, -\partial_x \psi)$. For a given flow field, the temperature field is obtained by solving
\begin{align}
\mathbf{u} \cdot \nabla T - \kappa \Delta T &= 0. \label{AdvDiff}
\end{align}
\nn The flow field is constrained to have fixed kinetic energy $KE$:
\begin{align}
\frac{1}{2}\rho W\iint |\mathbf{u}|^2 dA = KE, \label{KE}
\end{align}
\nn where $\rho$ is the fluid density and 
$W$ is the domain width in the out-of-plane ($z$) direction, along which all quantities
are assumed to be invariant. The flow field is not
explicitly constrained to satisfy the Navier-Stokes equations. However,
any (sufficiently smooth) incompressible flow satisfies the incompressible Navier-Stokes
equations with a suitable volume forcing term $\mathbf{f}$:
\begin{align}
\frac{D\mathbf{u}}{Dt} = -\frac{1}{\rho}\nabla p + \nu \Delta \mathbf{u} + \frac{1}{\rho}\mathbf{f} 
\quad ; \quad \nabla \cdot \mathbf{u} = 0. \label{NS}
\end{align}
\nn Given $\mathbf{u}$, (\ref{NS}) determines $\mathbf{f}$ and $p$. 
The flow can be considered to be driven by $\mathbf{f}$.

We maximize the (steady) rate of heat flux out of the hot boundary:
\begin{align}
Q = -\int_{r = 1} k\, \partial_n T ds. \label{Cond}
\end{align}
\nn Here $k$ is the thermal conductivity of the fluid, $n$ is the coordinate normal to the boundary, increasing into the fluid domain, and $s$ 
is the arc length coordinate along the boundary, 
here just the negative of the angular coordinate along the boundary, increasing from 0 to $2\pi$ moving along the circle in the clockwise direction.

Nondimensionalizing lengths by a typical length scale $L$ (in the present case, the radius of the inner cylinder), 
and time by a diffusion time scale $L^2/\kappa$, and writing $\mathbf{u}$ in terms of $\psi$, (\ref{AdvDiff}) becomes
\begin{align}
\partial_y \psi \partial_x T - \partial_x \psi \partial_y T - \Delta T &= 0, \label{AdvDiffND}
\end{align}
\nn and (\ref{KE}) becomes
\begin{align}
\iint |\nabla \psi|^2 dA = Pe^2, \label{Pe}
\end{align}
\nn where the Peclet number $Pe = \sqrt{2 KE/\rho W \kappa^2}$ measures the strength of advection relative to diffusion of heat, and is different from
the definition of \cite{hassanzadeh2014wall} (which uses the average flow speed) because we will deal with unbounded
fluid domains. We have already
nondimensionalized temperature by the temperature of the hot boundary 
(also the temperature difference between the boundaries). Having chosen scales for
length, time and temperature, we need to choose a typical mass scale to nondimensionalize (\ref{Cond}). Since mass enters the thermal conductivity,
for simplicity we instead chose a thermal conductivity scale to be that 
of the fluid, so that in dimensionless form, (\ref{Cond}) becomes
\begin{align}
Q = -\int_{r = 1} \partial_n T ds. \label{Conda}
\end{align}
\nn We maximize $Q$ over all steady 2D incompressible flow fields (given by $\psi$) of a given kinetic energy 
by taking the variation of the Lagrangian
\begin{align}
 \mathcal{L} =  -\int_{r = 1} \partial_n T ds + \iint m(x,y) \left(-\nabla^\perp \psi \cdot \nabla T - \Delta T\right) dA + 
\lambda \left(\iint |\nabla \psi|^2 dA - Pe^2 \right). \label{L}
\end{align}
\nn Here $m$ and $\lambda$ are Lagrange multipliers enforcing (\ref{AdvDiffND}) and (\ref{Pe}) respectively. The area integrals
are over the fluid domain, the annulus in figure \ref{fig:Annulus}A. The optimal $\psi$ is found
by setting to zero the variations of $\mathcal{L}$ with respect to $T$, $\psi$, $m$, and $\lambda$. Taking the variations and
integrating by parts, we obtain the following equations and boundary conditions:
\begin{align}
0 = \frac{\delta\mathcal{L}}{\delta m} &= -\nabla^\perp \psi \cdot \nabla T - \Delta T\quad , \quad T\;  \mbox{given on boundaries} \label{T} \\
0 = \frac{\delta\mathcal{L}}{\delta T} &= \nabla^\perp \psi \cdot \nabla m - \Delta m\quad , \quad \mbox{$m = T$ on boundaries} \label{m} \\
0 = \frac{\delta\mathcal{L}}{\delta \psi} &= -\nabla^\perp T \cdot \nabla m - 2\lambda\Delta \psi\quad , \quad \psi = \mbox{const. on boundaries} \label{psi} \\
0 = \frac{\delta\mathcal{L}}{\delta \lambda} &= \iint |\nabla \psi|^2 dA - Pe^2. \label{mu}
\end{align}
\nn The boundary conditions for (\ref{psi}) are due to the fact that the boundaries are solid walls with no flow penetration. 
There is slip along the walls, and the flow may be taken as an approximation of a flow which would occur with no-slip
boundary conditions. No-slip boundary conditions arise naturally when we consider the
problem with fixed enstrophy, later in this work. The constant
for $\psi$ may be taken to be zero on one boundary (the inner cylinder), to remove
arbitrariness in $\psi$. 
On other boundaries (here, the outer cylinder), the constant values of $\psi$ are unknowns set by the equations
\begin{align}
\int m \frac{\partial T}{\partial s} + 2 \lambda \frac{\partial \psi}{\partial n} ds = 0
\end{align}
\nn on each such boundary. Our approach is essentially the same as that of
\cite{hassanzadeh2014wall} so far. To solve the problem in this annular
geometry, and in a general class of geometries, it is convenient to change 
coordinates. Let $T_0$ denote the temperature with the given boundary
conditions and no flow, determined purely by conduction: $\Delta T_0 = 0$. 
$T_0$ is a harmonic function, here given by $1 - \log{r}/\log{R}$. 
It has a harmonic
conjugate function $\beta = -\theta/\log{R}$ which is related to
$T_0$ by the Cauchy-Riemann equations: $\partial_x T_0 = \partial_y \beta$,
$\partial_y T_0 = -\partial_x \beta$. Because the
flow domain is doubly-connected, $\beta$ is multi-valued; for simply
connected domains, a single-valued harmonic conjugate always exists
(\cite{brown1996complex}). $\beta$ is unique up to an additive constant.
$(T_0, \beta)$ are conformal coordinates,
and we will show that the equations (\ref{T})--(\ref{mu}) are essentially unchanged in these
coordinates. However, the flow domain transforms into a rectangle, and we will
show that this maps the problem to the one solved by \cite{hassanzadeh2014wall}.
For notational convenience, we set $\alpha(x,y) \equiv T_0(x,y)$ and use $\alpha$
as the coordinate name. The metric which gives the density of $(\alpha, \beta)$ equicontours in the $(x,y)$-plane is 
\begin{align}
h = \|\partial\mathbf{X}/\partial\alpha\| = \|\partial\mathbf{X}/\partial\beta\|, \quad\mbox{with}
\quad\mathbf{X} = (x(\alpha, \beta), y(\alpha, \beta)). \label{h}
\end{align}
\nn Writing $\nabla$, $\nabla^\perp$, and $\Delta$ in 
$(\alpha, \beta)$ coordinates using standard formulas for differential
operators in orthogonal
curvilinear coordinates (\cite{acheson1990elementary}) we have
\begin{align}
-\nabla_{x,y}^\perp \psi \cdot \nabla_{x,y} T - \Delta_{x,y} T = 
\frac{1}{h^2}\left(-\nabla_{\alpha,\beta}^\perp \psi \cdot \nabla_{\alpha,\beta} T - \Delta_{\alpha,\beta} T\right).
\end{align}
\nn Therefore equations (\ref{T})--(\ref{psi}) are unchanged in
$(\alpha, \beta)$ coordinates and so is (\ref{mu}) when the integral is
written in $(\alpha, \beta)$ coordinates:
\begin{align}
0 = \iint |\nabla_{\alpha,\beta} \psi|^2 dA_{\alpha,\beta} - Pe^2. \label{mua}
\end{align}
\nn The boundary conditions are also unchanged, except that they are
given on the sides of the rectangle (figure \ref{fig:Annulus}B). Thus
the problem is essentially the same as that solved by
\cite{hassanzadeh2014wall} with $(\alpha, \beta)$ here corresponding
to $(1-z, x)$ there. Instead of maximizing (\ref{Conda}) they maximized
the convective heat flux integrated over the flow domain. In Appendix
\ref{QAlt} we show that the two are equal in curvilinear coordinates, so
that in place of (\ref{Conda}) we can use
\begin{align}
Q =  \Delta \beta - \int_{\beta_{min}}^{\beta_{max}} \int_0^1 \partial_\beta \psi\, T \,d\alpha d\beta. \label{Q2}
\end{align}
\nn where $\Delta\beta \equiv \beta_{max}-\beta_{min}$ is the extent of the domain in $\beta$. 
Equation (\ref{Q2}) allows one to compute
the leading contributions to $Q$ in the small-$Pe$ limit
from the solution to a single eigenvalue problem. 

%
%

In the limit of small $Pe$, the solutions can be written as asymptotic series
in powers of $Pe$. When $Pe = 0$, the solutions to (\ref{T})--(\ref{mu}) are 
\begin{align}
m_0 = T_0 = \alpha = 1 - \frac{\log r}{\log R} \quad , \quad \psi_0 = 0 \quad , \quad \lambda \;  \mbox{undetermined}.
\end{align}
\nn Asymptotic solutions to (\ref{T})--(\ref{mu}) for $0 < Pe \ll 1$ are therefore posed as
\begin{align}
T &= T_0 + Pe \, T_1 + O(Pe^2) \label{Texpn} \\
m &= m_0 + Pe\, m_1 + O(Pe^2) \label{mexpn} \\
\psi &= Pe \,\psi_1 + O(Pe^2) \label{Psiexpn}
\end{align}
\nn Expanding (\ref{T})--(\ref{mu}) to $O(Pe)$ yields linearized equations for $T_1, m_1,\psi_1$:
\begin{align}
0 &= -\nabla^\perp \psi_1 \cdot \nabla T_0 - \Delta T_1, \quad T_1 = 0\;  \mbox{on boundaries} \label{T1} \\
0 &= \nabla^\perp \psi_1 \cdot \nabla m_0 - \Delta m_1, \quad m_1 = 0\;  \mbox{on boundaries} \label{m1} \\
0 &= -\nabla^\perp T_0 \cdot \nabla m_1 -\nabla^\perp T_1 \cdot \nabla m_0 - 2\lambda\Delta \psi_1\quad , \quad \psi_1 = \mbox{const. on boundaries} \label{psi1} \\
0 &= \iint |\nabla \psi_1|^2 dA - 1. \label{mu1}
\end{align}
\nn Examining (\ref{T1}) and (\ref{m1}) and using $m_0 = T_0$ we find
that $m_1 = -T_1$, so we can eliminate $m_0$ and $m_1$ and reduce (\ref{T1})--(\ref{mu1}) to
\begin{align}
0 &= -\nabla^\perp \psi_1 \cdot \nabla T_0 - \Delta T_1, \quad T_1 = 0\;  \mbox{on boundaries} \label{T1a} \\
0 &= -\nabla^\perp T_1 \cdot \nabla T_0 - \lambda\Delta \psi_1\quad , \quad \psi_1 = \mbox{const. on boundaries} \label{psi1a} \\
0 &= \iint |\nabla \psi_1|^2 dA - 1. \label{mu1a}
\end{align}
\nn Using $\nabla T_0 = \hat{\mathbf{e}}_\alpha/h$, the system simplifies to
\begin{align}
0 &= \partial_\beta \psi_1 - \Delta_{\alpha,\beta} T_1, \quad T_1 = 0\;  \mbox{on boundaries} \label{T1b} \\
0 &= \partial_\beta T_1 - \lambda\Delta_{\alpha,\beta} \psi_1\quad , \quad \psi_1 = \mbox{const. on boundaries} \label{psi1b} \\
0 &= \iint |\nabla_{\alpha,\beta} \psi_1|^2 d\alpha d\beta - 1. \label{mu1b}
\end{align}
\nn Because the boundary
conditions are periodic in $\beta$, the eigenfunctions may be found by taking a Fourier transform of 
(\ref{T1b})--(\ref{psi1b}) in $\beta$ (or $x$ in \cite{hassanzadeh2014wall}). We obtain
\begin{align}
\psi_1 &= A_{kn} \sin(k\pi\alpha)\sin(2\pi n\beta/\Delta\beta) \label{psi1def} \\
T_1 &= B_{kn} \sin(k\pi\alpha)\cos(2\pi n\beta/\Delta\beta)  \label{T1def}
\end{align}
\noindent where $\Delta\beta = 2\pi/\log{R}$.
We may add an arbitrary constant phase to $\beta$ which simply rotates the solutions without
changing the total kinetic energy or the heat transferred. 
The flow corresponding to (\ref{psi1def}) is an array of convection rolls in $(\alpha, \beta)$
space.
Inserting (\ref{psi1def})--(\ref{T1def}) into (\ref{T1b})--(\ref{psi1b}) we obtain $A_{kn}$, $B_{kn}$, and $\lambda$:
\begin{align}
A_{kn} &= \left( \frac{4/\Delta\beta}{(\pi k)^2 + (2\pi n/\Delta\beta)^2} \right)^{1/2},\\
B_{kn} &= \frac{-(4/\Delta\beta)^{1/2}(2\pi n/\Delta\beta)}
{\left[(\pi k)^2 + (2\pi n/\Delta\beta)^2\right]^{3/2}},\\
\lambda &= \frac{-(2\pi n/\Delta\beta)^2}
{\left[(\pi k)^2 + (2\pi n/\Delta\beta)^2\right]^{2}}.
\end{align}
\noindent The heat transferred by each mode (\ref{Q2}) can
be expanded in orders of $Pe$. The zeroth order term
is $Q_0 = \Delta \beta$. The first order term $Q_1$ 
vanishes for each mode (\ref{psi1def})--(\ref{T1def}), 
so the leading-order term to be maximized is
\begin{align}
Q_2 = - \int_{\beta_{min}}^{\beta_{max}} \int_0^1 \partial_\beta \psi_1\, T_1 \,d\alpha d\beta 
= -\lambda = \frac{(2\pi n/\Delta\beta)^2}
{\left[(\pi k)^2 + (2\pi n/\Delta\beta)^2\right]^{2}}. 
\end{align}

\begin{figure}
  \centerline{\includegraphics[width=14cm]
  {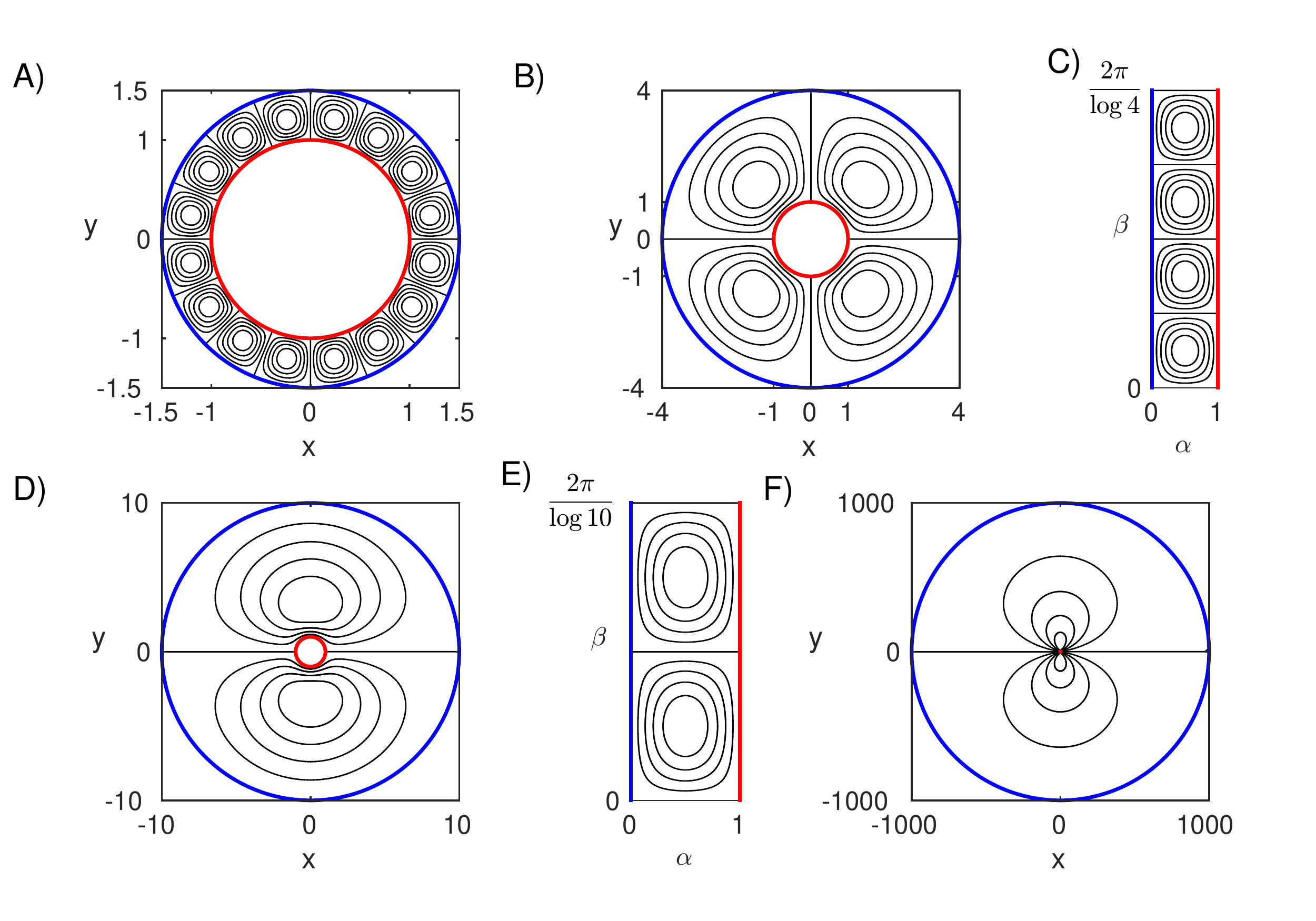}}
  \caption{Streamlines for optimal convection cooling flows in annular domains, 
with the outer (cold) boundary
at different radii $R$ = 1.5 (A), 4 (B), 10 (D), and 1000 (F). The inner (hot) boundary has radius 1. 
Panels C and E show the flows in panels B and D in $(\alpha, \beta)$ coordinates. The streamlines are plotted without arrows because there is no change in the heat transferred
when the flow direction is reversed. The streamlines are plotted at nine equally spaced values 
between the extrema of the stream function.
  }
\label{fig:AnnulusOptimaFig}
\end{figure}

\nn $Q_2$ is maximum when $k = 1$ and if $\Delta\beta/2$ is an integer, when $n = \Delta\beta/2$.
In this case, $Q_{2, max} = 1/4\pi^2$. If $\Delta\beta/2$ is not an integer but lies between the integers $l$ and $l+1$,
then the $Q_2$-maximizing $n$ is $l$ if $\Delta\beta/2 < \sqrt{l(l+1)}$ and $l+1$ otherwise. 
The $Q_2$-maximizing flow is a mode (\ref{psi1def}) consisting of approximately
square convection rolls in $(\alpha, \beta)$ space. In figure \ref{fig:AnnulusOptimaFig} we plot the streamlines at equal
intervals of the streamfunction for
$Q_2$-maximizing flows when the outer boundary radius is $R$ = 1.5, 4, 10, and 1000.

When $R$ is close to 1 (panel A), we have a thin gap between the hot and cold surfaces, and the flows 
tend to the square convection rolls in a straight channel found by \cite{hassanzadeh2014wall}. Since
$h$ in (\ref{h}) is $r \log{R}$, which varies little in the thin gap, there is relatively little distortion when the vortices are mapped
from the $(\alpha, \beta)$-plane to the $(x,y)$-plane.
The case $R \gg 1$
corresponds to the application of the convection cooling of the exterior of a circular body.
In this limit (e.g. panel F), the optimal flow is
\begin{align}
\psi_{max} = -A_{11} \sin{\left(\pi\left(1-\frac{\log{r}}{\log{R}}\right)\right)}\sin(\theta). \label{psi1opt}
\end{align}
\nn which consists of a pair of oppositely-signed vortices forming a dipole near the body. 
The amount of vortex distortion is given by $h = r\log{R}$, so the vortices are 
strongly stretched moving outward from the body. 
The centers of
the vortices are located at $r = \sqrt{R}$, the local extrema of $\psi_{max}$. Differentiating
(\ref{psi1opt}) with respect to $r$, we find that the azimuthal flow speed $u_\theta$ is proportional to
$1/r$ in the neighborhood of the body ($1 \leq r \ll R$). When
$\Delta\beta/2 = \pi/\log{R}$ is an integer, $Q_{2, max} = 1/4\pi^2$. For large
$R$, $Q_{2, max}$ is attained with $n = k = 1$.
Consequently, with a fixed total kinetic energy, the
optimal heat transferred by convection ($Pe^2 Q_{2, max} \sim Pe^2\log^{-2}{R}$) 
decays slowly as the outer radius increases. Physically, it seems reasonable
that the optimal heat transferred with fixed kinetic energy
should decrease as the hot and cold surfaces become more distant.

We have discussed optimal convection cooling using the specific example of an annular geometry, but
similar results can be obtained for a wide range of geometries. The first step is to
compute $(\alpha, \beta)$ for a given geometry. In the next two sections we discuss
the solutions for two geometries which are relevant to applications. We find that a slight
modification is needed in cases where the hot and cold surfaces meet.

\section{Hot plate embedded in a cold surface}\label{sec:plate}

\begin{figure}
  \centerline{\includegraphics[width=15cm]
  {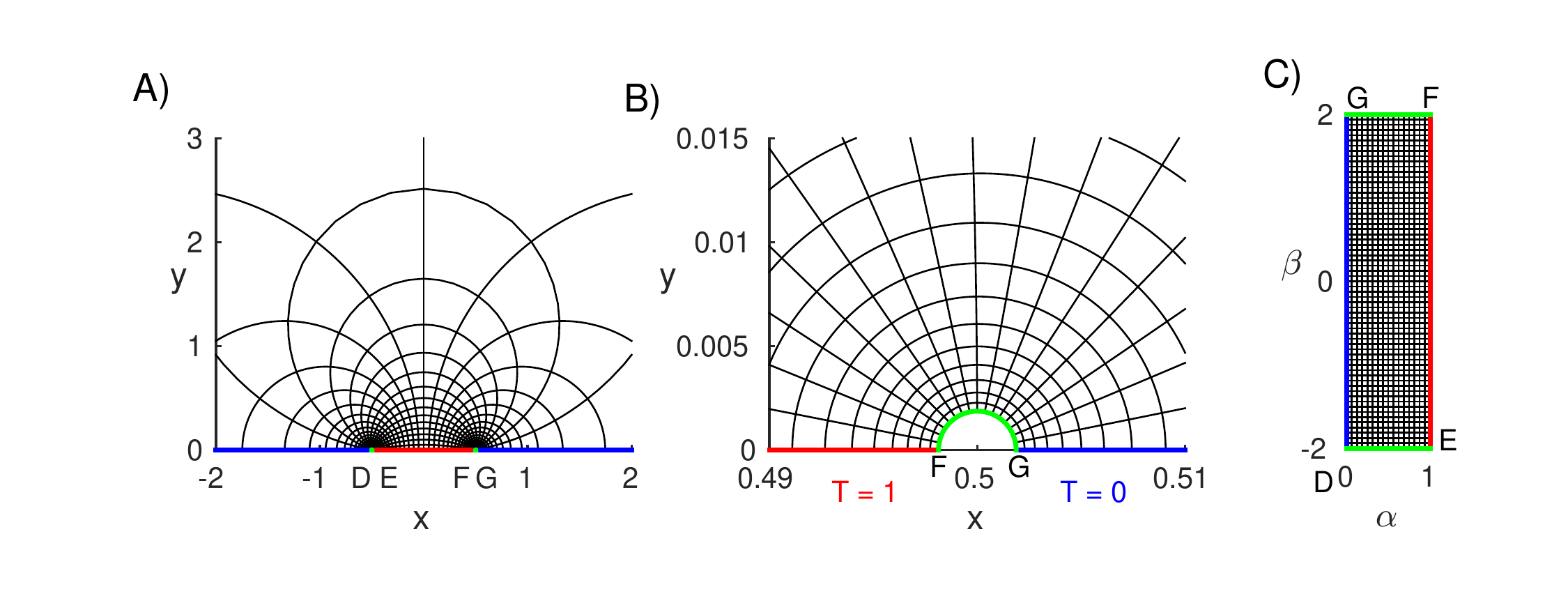}}
  \caption{A) A portion of the flow domain--the upper half 
plane--near the hot boundary. The hot boundary extends nearly to
the limits of the interval ($|x| \leq 1/2, y = 0$). The 
cold boundaries are approximately ($|x| > 1/2, y = 0$). They are joined
by small insulating boundaries which are approximately semi-circular.
The curvilinear grid plots equicontours of the conformal coordinates  
$\alpha = \frac{1}{\pi}\left(\arg{\left(z-\frac{1}{2}\right)} 
-\arg{\left(z+\frac{1}{2}\right)} \right)$ and $\beta = -\frac{1}{\pi}\log{\left|\frac{z-1/2}{z+1/2}\right|}$.
B) Close-up near $z = 1/2$. The approximate semi-circle closest to 1/2 (light grey or green) 
is the contour $\beta = 2$. This
is one of the insulating boundaries of the truncated domain, with the other
located near $z = -1/2$.  
C) The flow domain in ($\alpha, \beta$) space. The letters D, E, F, G show corresponding
points in panels A) and B).
  }
\label{fig:PlateFig}
\end{figure}

The next geometry we consider is a hot plate embedded in a cold surface 
(see figure \ref{fig:PlateFig}).  The hot plate gives a simple model of a flat heated surface such as a computer processor. 
An important difference with the previous case is that the cold and hot surfaces are now connected (or nearly so, as we will
discuss).  We nondimensionalize by the
length of the plate. 

The plate (where $T = 1$) occupies $-1/2 \leq x \leq 1/2$ on the real axis, and the cold surface (where $T = 0$) is $|x| > 1/2$. The fluid lies in the upper half plane $y > 0$.
In terms of
the complex coordinate $z = x+iy$, the pure conduction solution (with $Pe$ = 0) is
\begin{align}
T_0 =  \frac{1}{\pi}\left(\arg{\left(z-\frac{1}{2}\right)} -\arg{\left(z+\frac{1}{2}\right)} \right).
\end{align}
\nn Following the same procedure as before,
\begin{align}
\alpha =  \frac{1}{\pi}\left(\arg{\left(z-\frac{1}{2}\right)} 
-\arg{\left(z+\frac{1}{2}\right)} \right), \quad \beta =  -\frac{1}{\pi}\log{\left|\frac{z-1/2}{z+1/2}\right|} 
\label{abplate}
\end{align}
\nn These are also known as bipolar coordinates, with foci at $x = \pm1/2$. In these coordinates, the flow domain is $0 < \alpha < 1$,
$-\infty < \beta < \infty$. The problem is no longer periodic in $\beta$, 
and new boundary
conditions are required at $\beta = \pm\infty$. To proceed, we first truncate
the domain to  $0 < \alpha < 1$,
$\beta_{min} < \beta < \beta_{max}$, introducing new boundaries at 
finite values $\beta_{min} < 0$ and $\beta_{max} > 0$. 
By (\ref{abplate}), the boundaries' distances from $z = \pm 1/2$ decay 
exponentially fast ($\sim e^{-|\beta_{min}|\pi}$, $e^{-|\beta_{max}|\pi}$)
as $\beta_{min}$ and $\beta_{max}$ grow in magnitude. Therefore with moderate
values of $\beta_{min}$ and $\beta_{max}$
we obtain a good approximation to the original non-truncated domain in 
the $(x,y)$ plane.

\begin{figure}
  \centerline{\includegraphics[width=15cm]
  {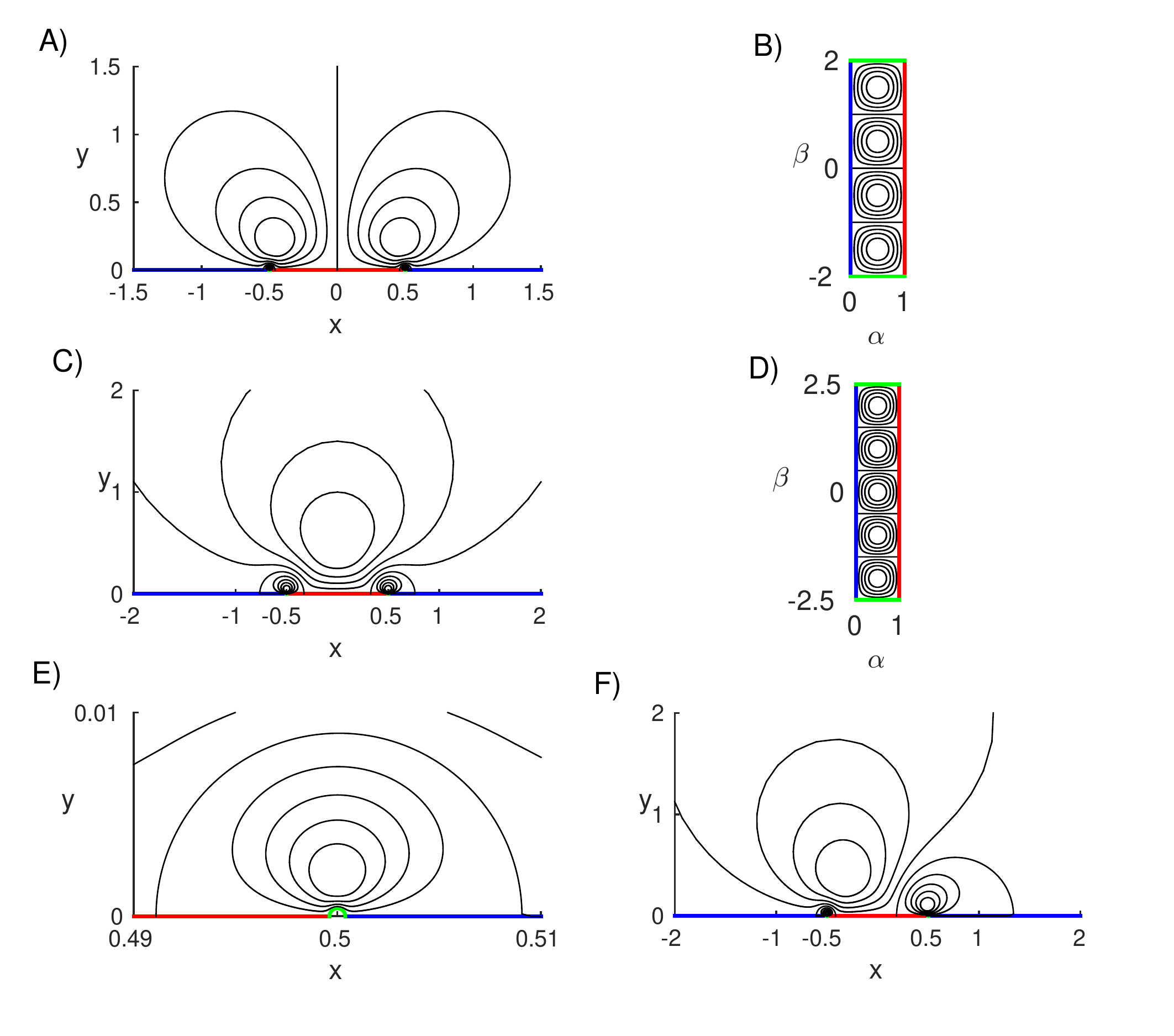}}
  \caption{A) Optimal heat transferring flow with $\beta_{min}$ = -2
and $\beta_{max}$ = 2. The flow has four vortices: one pair of large
vortices of the order of the plate length, and one pair of small vortices of the order of the insulating
boundary length. 
B) Flow from panel A in $(\alpha, \beta)$ coordinates. 
C) Optimal heat transferring flow with $\beta_{min}$ = -2.5
and $\beta_{max}$ = 2.5. The flow has five vortices, one
large vortex centered over the plate and two pairs symmetric about the y-axis.
D) Flow from panel C in $(\alpha, \beta)$ coordinates. 
E) Close-up of the flow in panel C near $z = 1/2$, showing one of the smaller vortices, near the insulated boundary (light gray or
green). F) Example of an asymmetric flow with $\beta_{min} = -2.75$, $\beta_{max} = 2.25$. Five
vortices are present, and the two nearest the insulated boundaries are too 
small to be visible. 
  }
\label{fig:PlateFlowsFig}
\end{figure}

We find
that the leading-order solution for $T$ is still $T_0$ if we make the new
boundaries insulating, so $0 = \partial_n T = \partial_\beta T/h$.  $T_0$
satisfies this boundary condition because $\partial_\beta T_0 = 
\partial_\beta \alpha = 0$ ($\alpha$ and $\beta$ are orthogonal coordinates).
Recomputing the variation of $\mathcal{L}$ with these Neumann (instead of periodic)
conditions at the $\beta$ boundaries,
we find that $m$ has the same boundary conditions as $T$, and thus as before,
$m_0 = T_0$, $m_1 = -T_1$. We assume the insulating boundaries are
solid walls that the flow does not penetrate. Because they
are connected to the hot and cold boundaries, this requires $\psi = 0$ on all four
sides of the rectangular domain in $(\alpha, \beta)$ space. Thus we obtain
the same eigenvalue problem as (\ref{T1b})--(\ref{mu1b}) but with
$\partial_n T_1 = \psi_1 = 0$ on the insulating
sides of the rectangle.
The eigenfunctions are slightly modified from (\ref{psi1def})--(\ref{T1def}):
 \begin{align}
\psi_1 &= A_{kn} \sin(k\pi\alpha)\sin(\pi n(\beta-\beta_{min})/\Delta\beta), \label{psi2def} \\
T_1 &= B_{kn} \sin(k\pi\alpha)\cos(\pi n(\beta-\beta_{min})/\Delta\beta)  \label{T2def}
\end{align}
\nn with
\begin{align}
A_{kn} &= \left( \frac{1/\Delta\beta}{(\pi k)^2 + (\pi n/\Delta\beta)^2} \right)^{1/2},\\
B_{kn} &= \frac{-(1/\Delta\beta)^{1/2}(\pi n/\Delta\beta)}
{\left[(\pi k)^2 + (\pi n/\Delta\beta)^2\right]^{3/2}},\\
\lambda &= \frac{-(\pi n/\Delta\beta)^2}
{\left[(\pi k)^2 + (\pi n/\Delta\beta)^2\right]^{2}}.
\end{align}
\nn Because the Neumann boundaries are solid insulating walls,
(\ref{Q2}) still holds, and $Q_2$, the leading contribution to the heat transferred by a given mode,
is 
\begin{align}
 Q_2 = -\lambda = \frac{ (\pi n/\Delta\beta)^2}
{\left[(\pi k)^2 + (\pi n/\Delta\beta)^2\right]^{2}}.
\end{align}
\nn $Q_2$ is maximized when $k = 1$ and $n$ is one of the two integers closest to
$\Delta\beta$, which again results in convection rolls which are square (or nearly so, if
$\Delta\beta$ is not an integer) in $(\alpha, \beta)$ space. In 
figure \ref{fig:PlateFlowsFig} we show examples 
of optimal flows when $\Delta\beta$ is even (panel A) and odd
(panels C and F). 

We have a chain of vortices
which are $O(1)$ in size above the hot plate, and shrink exponentially
in size as they approach the insulated boundaries at the hot-cold interface. In
panel A, $\beta_{min}$ = -2
and $\beta_{max}$ = 2, and there are four vortices, one large pair and one
small pair. The smallest
vortices are proportional in size to the insulated boundaries. Panel B shows the
same flow in the $(\alpha, \beta)$ plane. Panel C shows the optimal flow with $\beta_{min}$ = -2.5
and $\beta_{max}$ = 2.5, so there are five vortices (the smallest pair is not
visible). Panel D shows this flow in the $(\alpha, \beta)$ plane. Panel E shows
the flow near one of the small vortices in panel C at the insulated boundary
near $z = 1/2$.  Panel F shows an asymmetric case, $\beta_{min}$ = -2.75
and $\beta_{max}$ = 2.25. Again there are five vortices, but now they are
asymmetric with respect to the plate center.
The size of each vortex is
proportional to its distance from the insulated boundary, and the
typical flow speed within each vortex is inversely proportional to its typical length. 
Therefore, as the vortices become smaller, the flow speed increases such that the total
kinetic energy of each vortex (proportional to its area times typical flow speed squared)
is constant.  

Once again, $Q_{2, max} = 1/4\pi^2$ (when
$\Delta\beta$ is an integer), so the optimal heat transferred by convection is essentially
the same when we vary the positions of the insulating boundaries. The optimal flow varies considerably with the positions of the insulating boundaries, however, because they set the vortices' positions. It is somewhat
surprising that the optimal heat transferred by convection does not increase as the hot and cold surfaces are brought together (for the annulus we found a slow decrease in the heat transferred, but only when the surfaces are very distant). However, the heat transferred by conduction ($\Delta\beta$) diverges
logarithmically as the distance between the hot and cold surfaces tends to zero. 

\section{Channel with hot interior and cold exterior}\label{sec:channel}

\begin{figure}
  \centerline{\includegraphics[width=11cm]
  {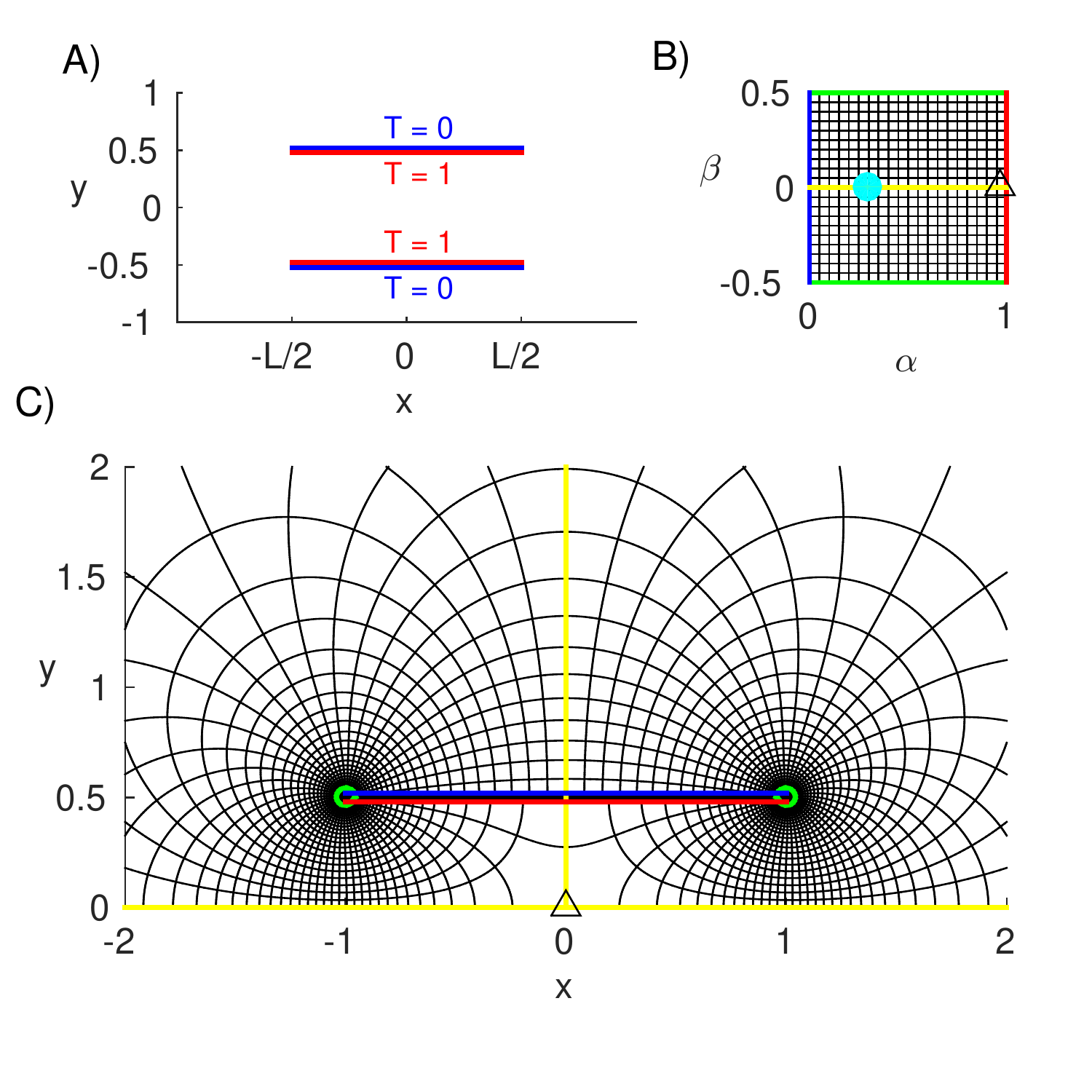}}
  \caption{A) Heated channel in an unbounded flow domain. The inside surfaces of the channel walls have temperature $T = 1$ and outside surfaces have $T = 0$.
B) The flow domain in $\alpha, \beta$ space for the case $L = 2, \beta_{min} = -0.5,$ and 
$\beta_{max} = 0.5$. The light blue (light gray) dot marks the point
at infinity and the black triangle marks the origin in $x,y$ space. 
The green boundaries are insulated surfaces, and the red and blue
boundaries are the hot and cold surfaces of the upper channel wall.
C) The upper half of the flow domain in $x,y$ space. The red, blue, green, and yellow 
lines and triangle correspond to those
in panel B. The curvilinear grid plots equicontours of $\alpha$ and $\beta$. 
  }
\label{fig:ChannelFig}
\end{figure}

A classic configuration for heat transfer is the flow through a heated
channel 
(\cite{eagle1930coefficient,dipprey1963heat,bird2007transport,lienhard2013heat,bejan2013convection}). 
Recent works have studied the dynamics of
flapping flags and vortex streets in 
channels (\cite{alben2015flag,wang2015dynamics}) with the goal of improving
heat transfer efficiency by using vortices to mix
the temperature field (\cite{gerty2008fluidic,shoele2014computational,jha2015small}). 
Here we consider a heated channel in an unbounded fluid (figure
\ref{fig:ChannelFig}A). 

The inside surfaces of the channel walls are
fixed at temperature $T = 1$ and the outside surfaces are fixed at
$T = 0$. If we assume that the channel flow is a 
perturbation of unidirectional flow, we could possibly truncate the
fluid domain to the interior of the channel, and use inflow and outflow
boundary conditions for the fluid and temperature 
(\cite{shoele2014computational,wang2015dynamics}). However, we wish to
make minimal assumptions about the flow and therefore we represent
the flow both inside and outside the channel. Energy is needed
to drive the flow through the entire fluid domain, so it makes sense to
include the outside flow in the optimization calculation.

The analytic function $\alpha + i\beta$ can be found numerically 
by using complex analysis and boundary integral methods. 
Because $T$ (and therefore $\alpha$)
has a jump across the channel walls, we can use the Plemelj formula
(\cite{estrada2012singular}) to write $\alpha + i\beta$ in
the form

\begin{align}
\alpha(z) + i\beta(z) = \frac{1}{2\pi i}\int_C \frac{\gamma(s) + i\eta(s)}{z - \zeta(s)} ds + C_1 \label{plemelj}
\end{align}
\nn where $C$ is a complex contour parametrized by arc length as
$\zeta(s)$, representing the union of the two channel walls, $s \pm i/2, -L/2 < s < L/2$. 
The real constant $C_1$ will be chosen shortly.
The function $\gamma(s) + i\eta(s)$ corresponds to the jump in $\alpha(s) + i\beta(s)$ along
the contour. 
From the boundary conditions on $\alpha$ (those of $T$) we have
$\gamma(s) = \pm 1$ on $\zeta(s) = s \pm i/2$. It remains to find $\eta(s)$. We
note that (\ref{plemelj}) provides the average value of $\alpha + i\beta$ on the
countour when $z$ is a point on the contour and the integral is changed from an
ordinary integral to a
principal value integral. We solve
for $\eta(s)$ by the condition that the average value of $\alpha(s)$ on the contour equals
$1/2$:
\begin{align}
1/2 = Re\left[\frac{1}{2\pi i}\barint_{-L/2}^{L/2} \left(\frac{1 + i\eta(s)}{s' - s} 
+ \frac{-1 + i\eta(s)}{s' - s +i}\right) \,ds \right] + C_1, \;\;-L/2 < s' < L/2. \label{nu}
\end{align}
\nn We have written the contributions from the two channel walls in (\ref{plemelj}) separately and
used symmetry considerations to deduce that $\eta$ on the lower wall equals
that on the upper wall. We use the Chebyshev collocation method (\cite{golberg1990ins})
to solve (\ref{nu}) with $s$ at points on a Chebyshev-Lobatto grid and $s'$ on a
Chebyshev-Gauss grid. We need two additional constraints to uniquely specify $\eta$ and $C_1$
in (\ref{nu}) (\cite{golberg1990ins}). These can be given as:
\begin{align}
\lim_{s \rightarrow \pm L/2} \eta(s) \sqrt{L^2/4-s^2} = 0.
\end{align}
\nn These conditions are preferred because they make $\eta$ a bounded continuous function. 
Having solved for $\eta$ and
$C_1$ and knowing $\gamma$, we evaluate $\alpha$ and $\beta$ using (\ref{plemelj}). As with the
single heated plate,
we find that $\beta \rightarrow \pm \infty$ at the plate edges where the hot and cold boundaries
meet. So as before, we cut off
the domain with small insulated boundaries at $\beta = \beta_{min}$ and $\beta_{max}$. In 
figure \ref{fig:ChannelFig}B and C we show $(\alpha,\beta)$ equicontours and their location in the $(x,y)$
plane, respectively, for the case $L = 2$. Here we use $\beta_{min} = -0.5, \beta_{max} = 0.5$. These
values are relatively small in magnitude so that the insulated boundaries are large enough to
be clearly visible. 
As with the single heated plate example, the insulated boundaries approach the plate 
edges exponentially with increasing
magnitude of these values.

In panel C, we show only the upper half plane in $(x,y)$. In the lower half plane,
the values of $\alpha$ and $\beta$ can be smoothly continued from the upper half plane, by reflecting the values in the origin. Thus the flow is also smoothly continued in
the lower half-plane by reflecting $\psi$ in the origin. 

\begin{figure}
  \centerline{\includegraphics[width=11cm]
  {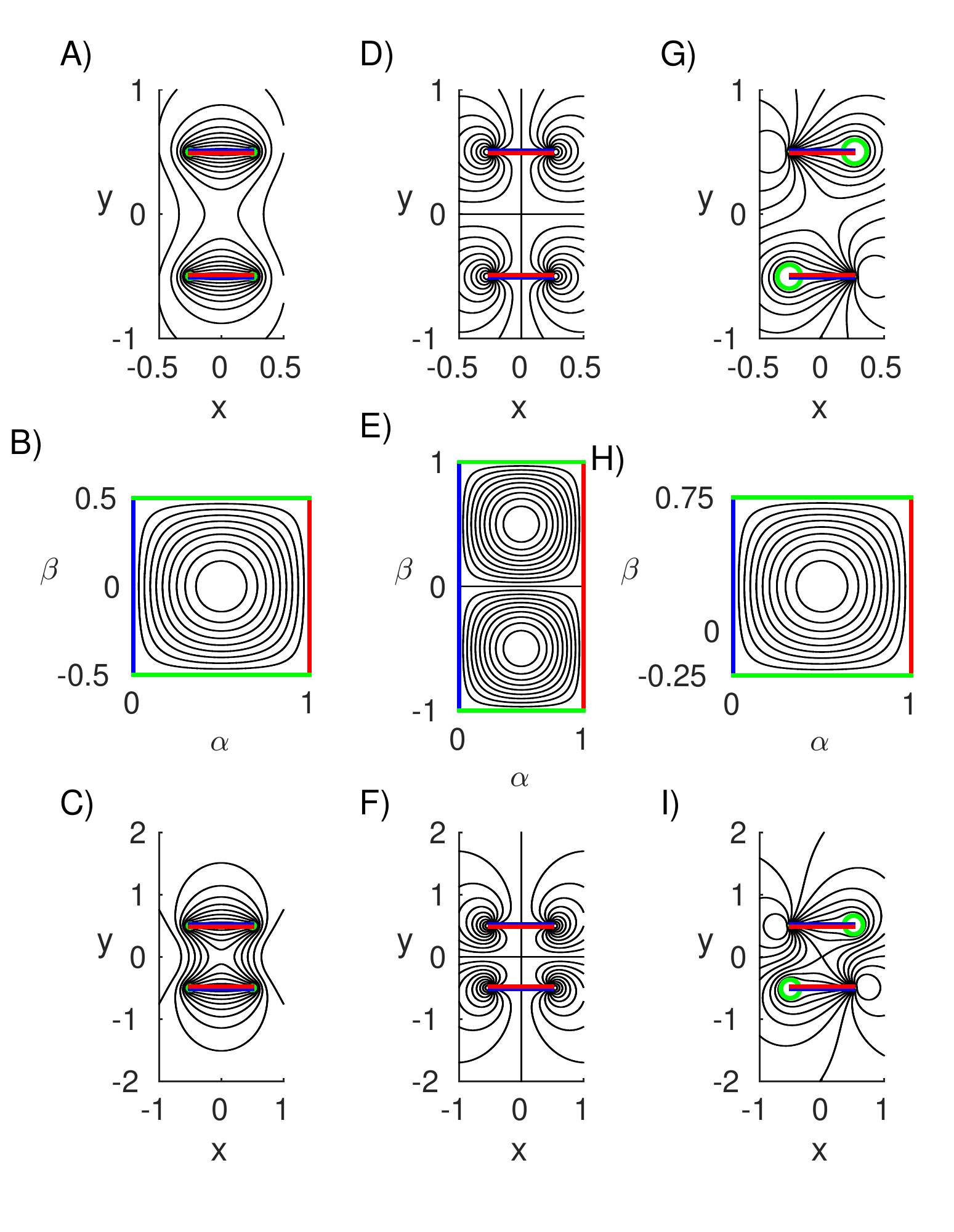}}
  \caption{Streamlines of optimal heat transferring flows for 
short channels. The channel lengths are $L = 0.5$ in the top
row (A, D, G) and 1 in the bottom row (D, F, I). The middle row (B, E, H)
shows the flows from the top and bottom rows in $(\alpha, \beta)$
coordinates, where they are the same. 
Moving from left to right, the positions of the insulated 
boundaries (green) varies: $\beta_{min} = -0.5,$ and 
$\beta_{max} = 0.5$ (A-C), $\beta_{min} = -1,$ and 
$\beta_{max} = 1$ (D-F, too close to the plate edges to be visible), and 
$\beta_{min} = -0.25,$ and $\beta_{max} = 0.75$ (G-I).}
\label{fig:ShortChannelFlowFig}
\end{figure}

In figure \ref{fig:ShortChannelFlowFig} we show optimal flows
for ``short'' channels, with lengths less than or equal to the
channel height. The top row (A, D, G) shows optimal flows when
the channel length is half the height. The only difference between
A, D, and G is the location of the insulating boundaries. As we
have seen with the single plate, 
this makes a big difference in the optimal flow. In panel A, the flow
consists of rolls that move around the plates. There is a strong net flow
through the channel, which is fast near the channel walls, and slow 
in the middle of the channel. The corresponding flow in the 
$(\alpha, \beta)$ plane is shown in panel B. It is a single vortex which
is mapped twice onto the $(x, y)$ plane in panel A by reflection
in the origin, as mentioned previously. In panel D we move the insulating
boundaries closer to the plate edges, and obtain two pairs of vortices moving
around the edges of the channel. The flow is fastest near the plate edges,
where the hot and cold surfaces are closest. Panel E shows the corresponding
flow in the $(\alpha, \beta)$ plane. Panel G shows an example
in which the insulating boundaries are asymmetric in $\beta$, with the 
$(\alpha, \beta)$ representation shown in panel H. 

The bottom row (C, F, I) shows optimal flows when the channel length
equals its height. The corresponding flows in the $(\alpha, \beta)$ plane
are again those in panels B, E, and H.
In panel C, the flow has a stronger circulatory component that moves
across the channel openings, from top to bottom and vice versa, 
around the outside of the entire channel, compared to that in panel A.
In panels F and I, the flow is also stronger near the centers of the
channel openings than in the corresponding flows of the top row (D and G).

\begin{figure}
  \centerline{\includegraphics[width=11cm]
  {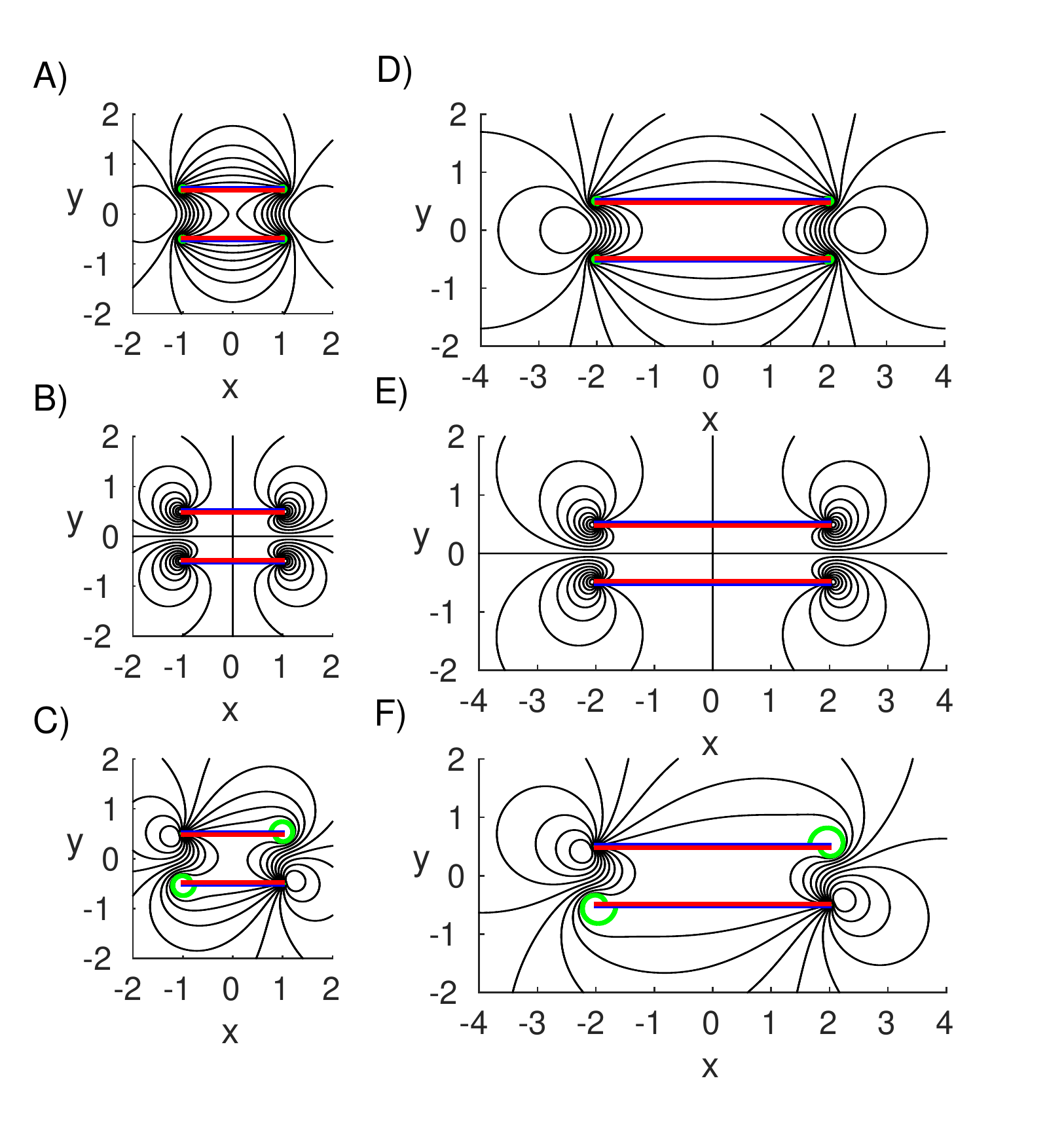}}
  \caption{Streamlines of optimal heat transferring flows for 
longer channels (lengths $L = 2$ (A-C)
and 4 (D-F) in units of channel width). 
For the length-2 channels (A-F), the insulated boundaries (green) 
are located at A) $\beta_{min} = -0.5,$ and 
$\beta_{max} = 0.5$, B) $\beta_{min} = -1,$ and 
$\beta_{max} = 1$ (too close to the plate edges to be visible), and 
C) $\beta_{min} = -0.25,$ and $\beta_{max} = 0.75$.
For length-4 channels (D-F), the insulated boundaries (green) 
are located at D) $\beta_{min} = -0.5,$ and 
$\beta_{max} = 0.5$, E) $\beta_{min} = -1,$ and 
$\beta_{max} = 1$ (too close to the plate edges to be visible), and 
F) $\beta_{min} = -0.25,$ and $\beta_{max} = 0.75$.
  }
\label{fig:LongChannelFlowFig}
\end{figure}

In figure \ref{fig:LongChannelFlowFig} we show optimal flows
when the channel length is increased to two (A-C) and four (D-F)
times the channel height. Now the flows are strongly confined to the
channel openings. There is very little flow in the center of the
channel. The reason is that the temperature due to pure
conduction ($T_0 = \alpha$) is nearly uniform in the center of
the channel. There is little to be gained by using energy to transport fluid through
this region of nearly uniform temperature. Since $\alpha$ changes little
in this region, $h = \|\partial\mathbf{X}/\partial\alpha\| = \|\partial\mathbf{X}/\partial\beta\| \gg 1$ and the flow speed
$\|\mathbf{u}\| = \|-\nabla^\perp_{x,y} \psi\| =\| -\nabla^\perp_{\alpha,\beta} \psi/h\|
\sim 1/h \ll 1.$ The lack of flow through the channel is a significant
difference from typical channel convection cooling flows 
(\cite{bird2007transport}). This can be attributed to our objective of maximizing the total heat flux out of the hot walls. In these solutions,
most of the flux
is close to the channel openings, near the cold exterior.

\section{Large amplitude flows}\label{sec:LargeKE}

We now consider the case where $Pe$ is not small by returning to equations 
(\ref{T})--(\ref{mu}), which hold for all $Pe \geq 0$.
We have noted that this is the same as the constant-kinetic-energy
problem considered by \cite{hassanzadeh2014wall} under the transformation
$(\alpha, \beta) \rightarrow (1-z, x)$. They computed numerically the solutions
from  $Pe$ small to large and derived a boundary layer solution for
the optimal flow in the limit of large $Pe$. Their derivation works 
for both the periodic and Neumann
boundary conditions (in $\beta$) that we used here. We simply translate
the complete asymptotic solution for $\psi$ that they found into an
$(\alpha, \beta)$ rectangle with $\beta$-width $\Delta\beta$:

\begin{align}
 \psi &\sim \frac{f(\mu_H,\Gamma)}{\sqrt{2\mu_H}}\sin\left(\frac{N\pi(\beta-\beta_{min})}{\Delta\beta}\right)
\tanh\left(\frac{\pi f(\mu_H,\Gamma)(1-\alpha)}{2\sqrt{2\mu_H}\,\Gamma}\right)
\tanh\left(\frac{\pi f(\mu_H,\Gamma)\alpha}{2\sqrt{2\mu_H}\,\Gamma}\right)
\nonumber \\
& \mbox{where} \quad f(\mu_H,\Gamma) = 1 - \frac{\pi\sqrt{2\mu_H}}{2\Gamma}. \label{largePe}
\end{align}
\nn In the limit of large $Pe$, equal-aspect-ratio convection rolls are no longer optimal.
Instead, the convection rolls become very narrow in the $\beta$ direction,
with narrow boundary layers of fast-moving fluid along the hot and cold walls.
The size of the boundary layers and the $\beta$-width of the rolls are set by the parameters
$\Gamma$ and $\mu_H$ (= $2\lambda$) in (\ref{largePe}). For $Pe \gg 1$ 
\cite{hassanzadeh2014wall} found that for the optimal rolls,
\begin{align}
\Gamma \sim 3.8476 Pe^{-1/2}\Delta\beta^{1/4} \quad ; \quad\mu_H = \Gamma^2/8\pi^2 ,
\end{align}
\nn where we have added the $\Delta\beta$ term due to our different definition of $Pe$.
This solution is valid for our problems with Neumann conditions at the
$\beta$ boundaries when $\Delta\beta/\Gamma$ is an
integer, $N$ in (\ref{largePe}). For periodic boundary conditions, $N$ must
be even. 

\begin{figure}
  \centerline{\includegraphics[width=15cm]
  {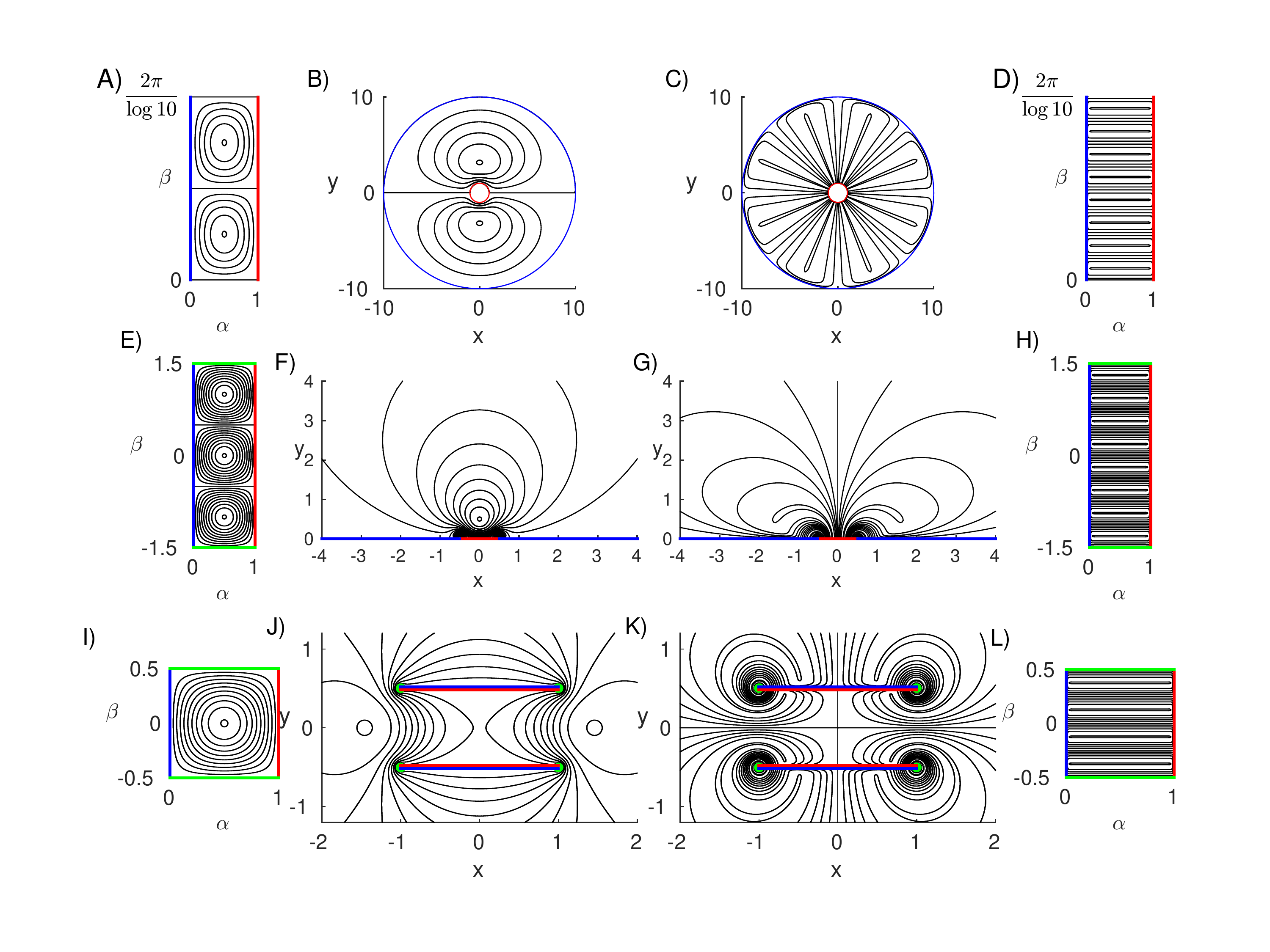}}
  \caption{A comparison of optimal flows with small and large kinetic energy
($Pe$) across the three geometries considered in this work. Top row (A-D): 
optimal flows in an annulus with outer radius $R = 10$. A) Small-$Pe$
optimal flow in $(\alpha, \beta)$. B) Small-$Pe$
optimal flow in $(x, y)$. C) Large-$Pe$ ($Pe \approx$ 210)
optimal flow in $(x, y)$. D) Large-$Pe$ ($Pe \approx$ 210)
optimal flow in $(\alpha, \beta)$.
Middle row (E-H): optimal flows over a hot plate in a cold surface.
E) Small-$Pe$
optimal flow in $(\alpha, \beta)$. F) Small-$Pe$
optimal flow in $(x, y)$. G) Large-$Pe$ ($Pe \approx$ 182)
optimal flow in $(x, y)$. H) Large-$Pe$ ($Pe \approx$ 182)
optimal flow in $(\alpha, \beta)$. 
Bottom row (I-L): optimal flows around a channel with hot interior
and cold exterior. I) Small-$Pe$
optimal flow in $(\alpha, \beta)$. J) Small-$Pe$
optimal flow in $(x, y)$. K) Large-$Pe$ ($Pe \approx$ 237)
optimal flow in $(x, y)$. L) Large-$Pe$ ($Pe \approx$ 237)
optimal flow in $(\alpha, \beta)$. 
  }
\label{fig:LargeAmplitudeFlows}
\end{figure}

In figure \ref{fig:LargeAmplitudeFlows} we plot a few examples of optimal flows at moderately large $Pe$ for the geometries we have considered, together with the corresponding
small-$Pe$ optima. The top row shows flows past a cylinder with outer boundary
at $R = 10$ at small $Pe$ in $(\alpha, \beta)$ (A) and $(x, y)$ (B), the same
flows as in figure \ref{fig:AnnulusOptimaFig}E and D, respectively. The large-$Pe$ optimum
is shown in panels C and D of figure \ref{fig:LargeAmplitudeFlows}. 
$Pe \approx 210$ so that $N = 8$. The streamlines are approximately rectangles
in $(\alpha, \beta)$ (D), which map onto approximate wedges in $(x, y)$ (C).
In the middle row, we show optimal flows over a hot plate
at small $Pe$ (E, F, similar to those in figure \ref{fig:PlateFlowsFig}) 
and $Pe \approx 182$ (G, H). A given rectangular streamline in $(\alpha, \beta)$ (H)
maps onto a streamline which approximately follows one larger circle and one smaller
circle (in the opposite direction) in the bipolar coordinate system (figure \ref{fig:PlateFig}A) connected by segments along the hot and cold plates. 
In the bottom row, optimal flows in a channel of length 2 are shown at small
$Pe$ (I, J) and $Pe \approx 237$ (K, L), giving $N = 4$. In this case a rectangular
streamline in $(\alpha, \beta)$ maps onto a streamline which follows two arcs
which are roughly circles centered at the channel edges. The arcs are again
connected by segments along the hot and cold plates. 

\section{More general geometries}\label{sec:General}

We have studied three examples which are representative of a more
general class of problems which can be solved with the same method. The main 
step is to solve for 
$(\alpha(x,y), \beta(x,y))$. The first case,
flow through an annulus, can be extended to any doubly connected region which
lies between one simple closed curve on which the temperature is 1 and another
on which the temperature is 0. The flow region may include the point at infinity.
All such regions correspond to a rectangle in $(\alpha, \beta)$ space on
which the boundary conditions for $\psi$ and $T$ are Dirichlet on the 
$\alpha = 0$, $\alpha = 1$
sides and periodic on the $\beta = \beta_{min}$, $\beta = \beta_{max}$
sides. 

The second case, flow over a heated plate, can be generalized to the flow within
any region bounded by a simple closed curve which is partitioned into four
arcs. On the four arcs the boundary
conditions are $T = 0$, $\partial_n T = 0$, $T = 1$, and $\partial_n T = 0$, moving
continuously around the curve. On the entire curve $\psi = 0$. The curve may
include the point at infinity (as for the heated plate), and in general we may
use any simple closed curve on the Riemann sphere. On the arcs where $\partial_n T = 0$,
we have $0 = \partial_n T_0 = \partial_n \alpha = -\partial_s \beta$, so $\beta$ is
constant. All such cases correspond
to a rectangle in $(\alpha, \beta)$ space on
which the boundary conditions are Dirichlet for $\psi$, and Dirichlet and Neumann for
$T$ on the $\alpha$ and $\beta$ sides, respectively. 
The green arcs in figures 3-7 were chosen because they followed lines of constant
$\beta$ for functions which could be written in closed form or computed relatively 
easily, but the shapes of the insulated boundaries can be arbitrary in general.

The third case, flow through a heated channel, is an example where the flow lies
in the region between two simple closed curves, each of which is partitioned into
four arcs as above. Here by symmetry the same rectangle in $(\alpha, \beta)$ was used
to cover the $(x,y)$ space twice. This case can be generalized to nonsymmetrical 
configurations of $n$ closed curves, and the curves may contain multiple arcs on which  
$T = 0$ and $T = 1$. 

\begin{figure}
  \centerline{\includegraphics[width=14cm]
  {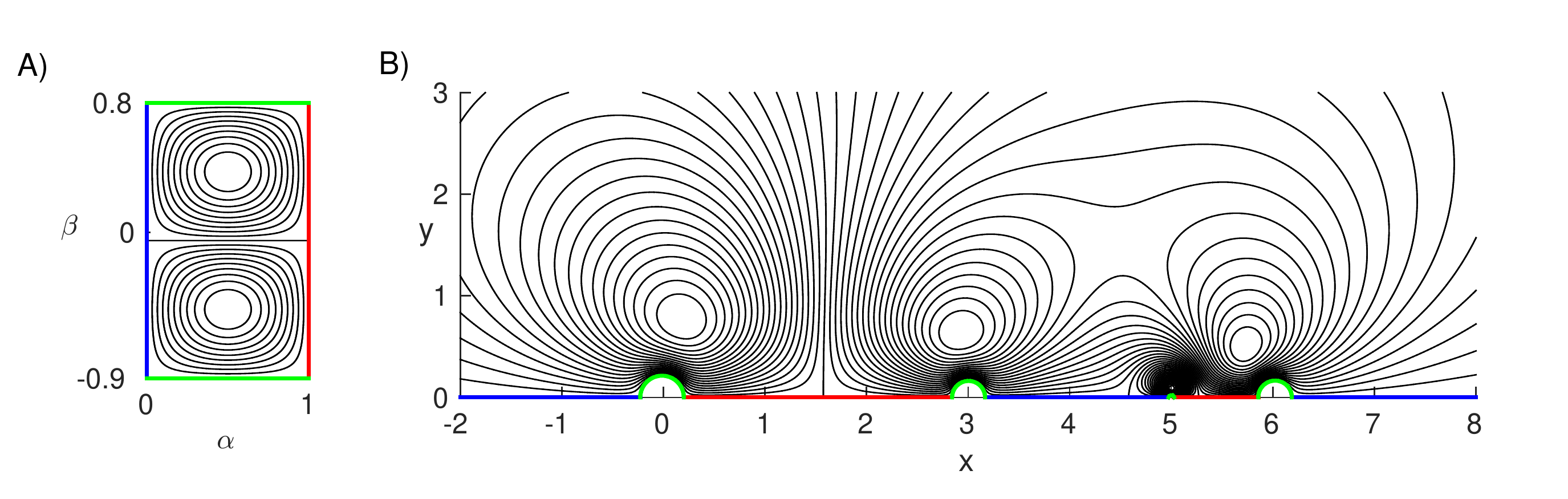}}
  \caption{Optimal small kinetic energy flow (small $Pe$) over a surface
with two hot plates (red) separated by cold plates (blue) and insulated
boundaries (green), in $(\alpha, \beta)$ (A)
and $(x,y)$ (B). The flow is described by (\ref{psi2def}) with
 $\beta_{min} = -0.9$, 
$\beta_{max} = 0.8$, and
($\alpha, \beta$) given by (\ref{multiplate}).
  }
\label{fig:ExoticRegion}
\end{figure}

Figure \ref{fig:ExoticRegion} shows an example where the flow domain is the upper
half plane, bounded by two hot and three cold segments along the
$x$ axis. On the Riemann sphere this is one closed curve on which
the temperature is 1 on two arcs and 0 on two arcs. Here
\begin{align}
\alpha =  \frac{1}{\pi}\arg{\left(\frac{(z-6)(z-3)}{(z-5)z}\right)}, 
\quad \beta =  -\frac{1}{\pi}\log{\left|\frac{(z-6)(z-3)}{(z-5)z}\right|}. 
\label{multiplate}
\end{align}
\nn The rectangle with sides $\alpha = 0$, $\alpha = 1$, $\beta = -0.9$, 
$\beta = 0.8$ maps twice onto the flow region. Moving around the
rectangle counterclockwise twice in panel A corresponds to moving along the boundary on
the $x$ axis and green arcs in panel B from $x = -\infty$ to $+\infty$.
In the small-$Pe$ limit
the stream function is (\ref{psi2def}) with $n = 2$.

\section{Fixed enstrophy flows}\label{sec:Enstrophy}

\cite{hassanzadeh2014wall} also considered the optimization problem
with fixed enstropy instead of fixed kinetic energy. Fixed
enstropy is equivalent to a fixed rate of viscous energy 
dissipation, which is proportional to
the enstrophy for our flow boundary conditions (solid walls or periodic boundaries)
(\cite{lamb1932hydrodynamics}). In place of
(\ref{Pe}) we have:
\begin{align}
\frac{1}{2}\iint \nabla \mathbf{u}:\nabla \mathbf{u} \, dA = 
\iint (\Delta \psi)^2 dA = Pe^2. \label{Enstrophy}
\end{align}
\nn Here $Pe = \sqrt{\dot{E}/\mu W}$ is redefined to 
include the (steady) total rate of viscous energy dissipation $\dot{E}$ instead of the total kinetic
energy, the fluid viscosity $\mu$, and the out-of-plane width $W$. 
With (\ref{Enstrophy}) instead of (\ref{Pe}) in the Lagrangian (\ref{L}),
the equation for $\psi$ becomes 
\begin{align}
0 = -\nabla^\perp T \cdot \nabla m - 2\lambda\Delta^2 \psi\quad , 
\quad \psi = \mbox{const.}, \; \partial_n \psi = 0 \; \mbox{on boundaries}. \label{psi2}
\end{align}
instead of (\ref{psi}); (\ref{psi2}) includes the biharmonic operator and therefore 
an additional boundary condition on $\psi$, the no-slip condition corresponding to a viscous flow. 
In $(\alpha,\beta)$ coordinates, (\ref{Enstrophy}) becomes
\begin{align}
\iint \frac{1}{h^2} (\Delta_{\alpha,\beta} \psi)^2 dA_{\alpha,\beta} = Pe^2. \label{EnstrophyAB}
\end{align}
\nn and (\ref{psi2}) becomes
\begin{align}
0 = -\nabla_{\alpha,\beta}^\perp T \cdot \nabla_{\alpha,\beta} m - 2\lambda\Delta_{\alpha,\beta}
\left(\frac{1}{h^2}\Delta_{\alpha,\beta} \psi \right) , 
\quad \psi = \mbox{const.}, \; \partial_n \psi = 0 \; \mbox{on boundaries}. \label{psi2ab}
\end{align}
\nn The metric term ($1/h^2$) in (\ref{psi2ab}) is a function of $(\alpha,\beta)$ 
given by (\ref{h}), 
different for each geometry, and therefore
unlike for the fixed kinetic energy case, the solutions are not geometry-independent
when viewed in $(\alpha,\beta)$ coordinates. 

The case of a straight channel between
hot and cold walls (where $h \equiv 1$) was solved by 
\cite{hassanzadeh2014wall} with free-slip
boundary conditions instead of no-slip conditions, to simplify the problem.
In this case, for small $Pe$ the optimal flows are sinuosoidal convection
rolls with the same form as for the fixed kinetic energy case, but with a
different aspect ratio ($\sqrt{2}$ versus 1). At large $Pe$, the boundary
layer structure is different, but the outer flow has the same form as
in the fixed kinetic energy case. In the no-slip case, the optimal
flows with fixed enstrophy were
computed numerically by \cite{souza2016optimal} and compared to the
free-slip case. The solutions are similar qualitatively in both
cases at small and large $Pe$. In both cases there is a transition from convection
rolls with ``round'' streamlines and a moderate aspect ratio at small $Pe$ to 
narrow rectangular convection
rolls at large $Pe$. At large $Pe$ 
there are significant differences in the boundary layer structure, 
and a small difference in the power-law scaling of heat transferred ($Q$)
with respect to $Pe$: $Q \sim Pe^{0.58}$ for the stress-free optima while
$Q \sim Pe^{0.54}$ for the no-slip optima. 

For more general geometries, the fixed-enstrophy problem requires a numerical
solution. The $(\alpha,\beta)$ coordinates are useful here because they map the
flow domain onto a rectangle, where the differential operators can be discretized
by finite differences on uniform grids. The ($1/h^2$) factor
in (\ref{EnstrophyAB}) can be viewed as a weight that decreases the cost of
vorticity where $h$ is large. By (\ref{h}), $h = \|\partial_x \alpha, 
\partial_y \alpha\|^{-1} = \| \nabla T_0 \|^{-1}$, so the cost of vorticity
is decreased where $\| \nabla T_0 \|$ is
small. Typically this occurs far from the hot surface.

\begin{figure}
  \centerline{\includegraphics[width=14cm]
  {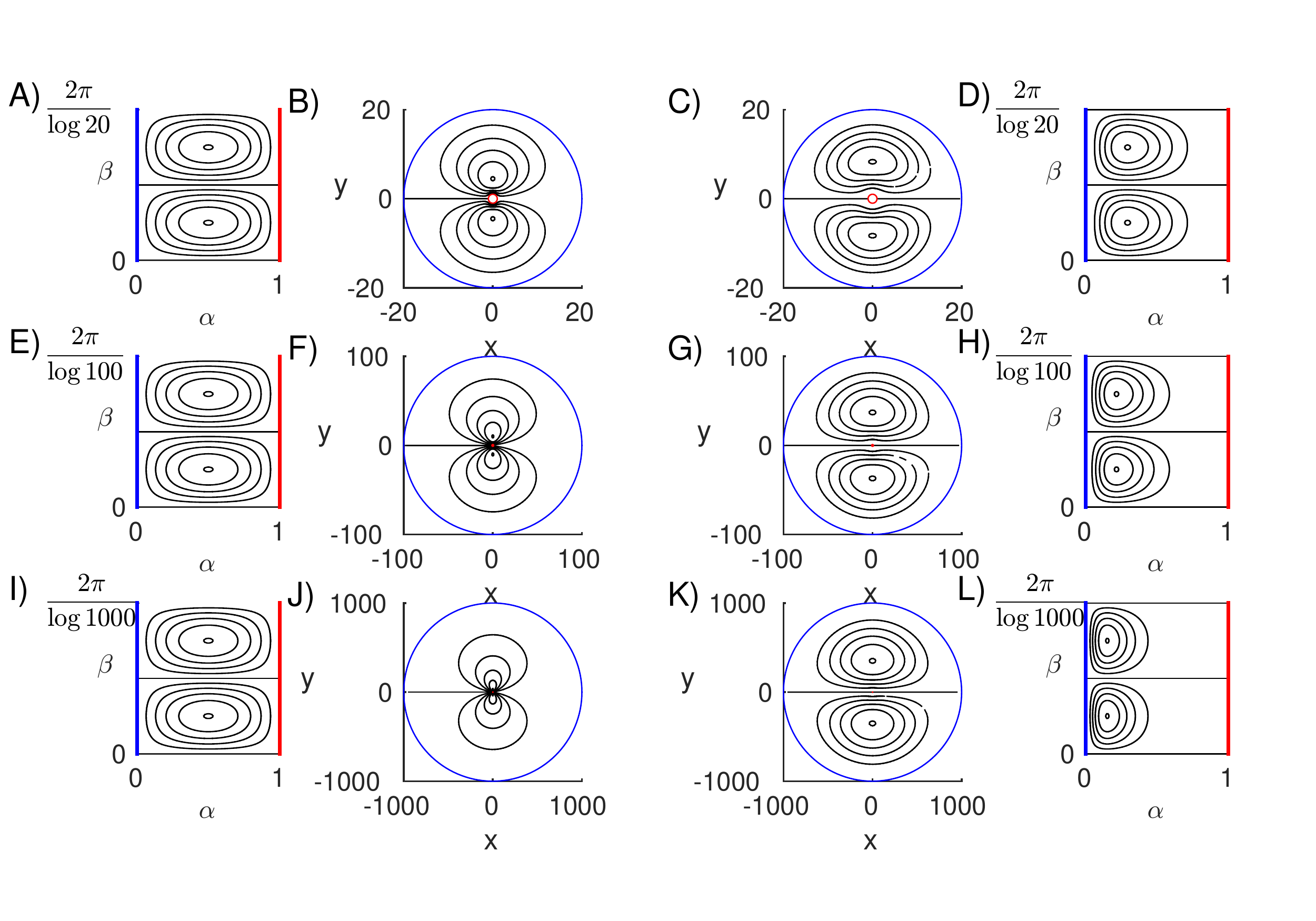}}
  \caption{Comparison of small-$Pe$ optimal flows in an annulus with fixed kinetic energy 
(left two columns) and fixed enstrophy (right two columns). The outer
boundary radius is $R$ = 20 (top row, A-D), 100 (middle row, E-H), and
1000 (bottom row, I-L). The fixed kinetic energy flows 
are shown in the first column (A, E, I) in the $(\alpha, \beta)$ plane
and in the second column (B, F, J) in the $(x,y)$ plane. The fixed-enstrophy
flows are shown in the third column (C, G, K) in the $(x,y)$ plane 
and in the fourth column (D, H, L) in the $(\alpha, \beta)$ plane.
  }
\label{fig:AnnulusEnstrophy}
\end{figure}

We now discuss the solutions to the small-$Pe$ eigenvalue 
problem with fixed enstrophy for the annular
domains with various outer radii $R$. In the third column of
figure \ref{fig:AnnulusEnstrophy} we present the optimal flows at 
$R$ = 20 (C), 100 (G), and 1000 (K) in the
$(x,y)$ plane. When lengths are scaled by $R$, the flows converge to a
common solution in the large-$R$ limit, consisting of a smooth vortex dipole
which is the size of the domain
(and obeys no-slip at the boundaries). In the fourth column, the same flows
are shown in the $(\alpha,\beta)$ plane. As $R$ increases (moving from panel D to
H to L), the vorticity moves towards $\alpha = 0$, 
where $h = R^{1-\alpha}\log(R)$ is largest. In the second column (B, F, J) we plot the
optimal flows with fixed kinetic energy. In these cases the vortex dipole
is stretched towards the inner cylinder, with each vortex center at a distance 
$\sim \sqrt{R}$ from the origin. In the first column, the same flows are shown in the
$(\alpha,\beta)$ plane, where they become identical under a rescaling
of the $\beta$ axis. 

With fixed kinetic energy, we have larger vorticity near
the inner cylinder. This vorticity is spread more uniformly in the 
fixed-enstrophy solutions. We now show that for these flows 
(i.e. panels C, G, and K), the kinetic energy diverges in the limit of large $R$. 
For the fixed-enstrophy flows, we have $\iint (\Delta_{x,y} \psi)^2 dA = Pe^2$. 
Thus $(\Delta_{x,y} \psi)^2 R^2 \sim Pe^2$ so $Pe/R \sim \Delta_{x,y} \psi \sim \psi/R^2$.
Thus $\psi \sim R$ and $\mathbf{u} = -\nabla_{x,y}^\perp \psi \sim \mathbf{1}$. Thus the flow
speed is of order 1 over an area $\sim R^2$, so the kinetic energy diverges as
$R^2$ as $R$ becomes large.

\begin{figure}
  \centerline{\includegraphics[width=14cm]
  {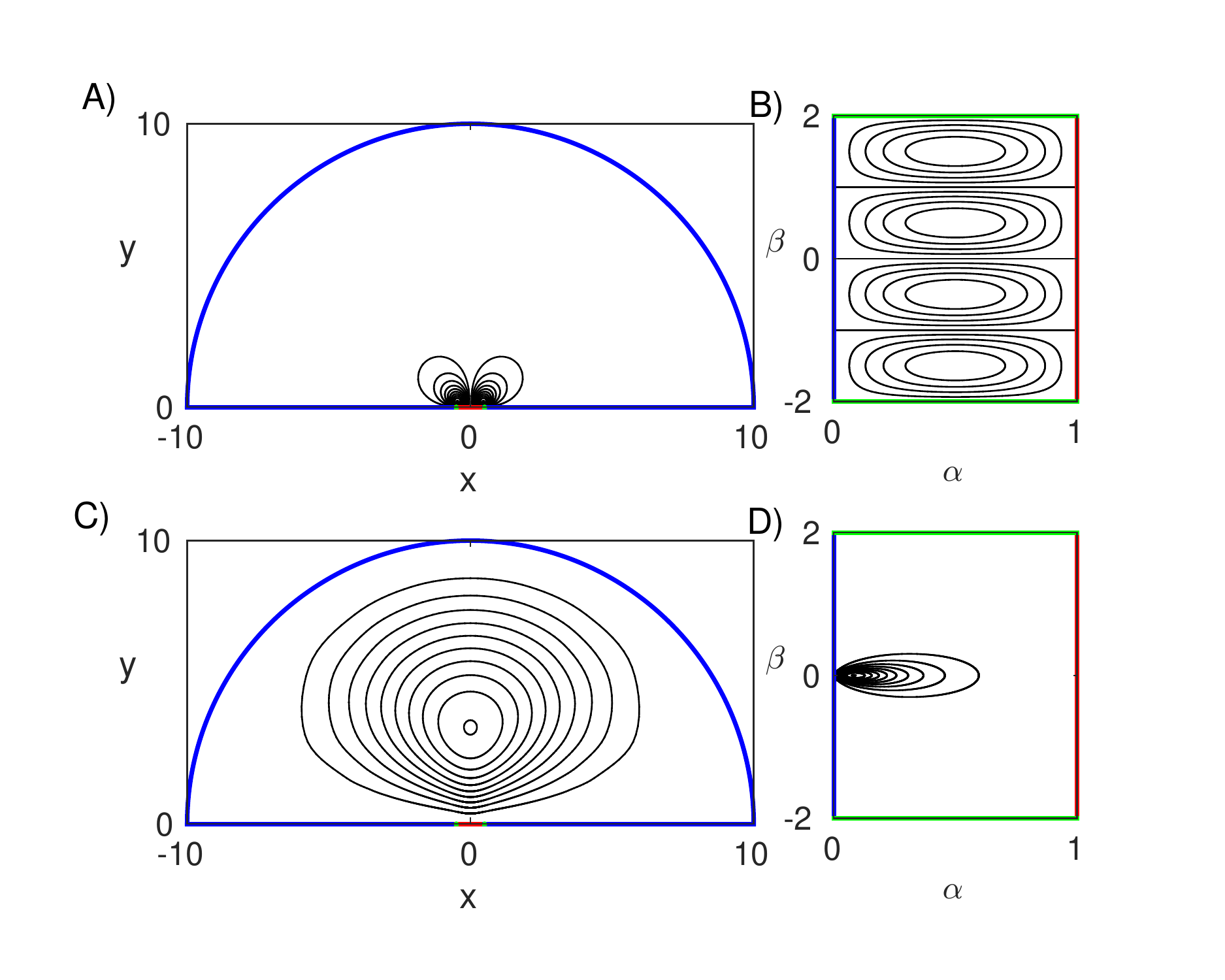}}
  \caption{Comparison of small-$Pe$ optimal flows with fixed kinetic energy 
(top row) and fixed enstrophy (bottom row) in a half-disk with radius
10 above a heated plate of unit length centered at the origin. Optimal
flow with fixed kinetic energy in the $(x,y)$ plane (A) and the
$(\alpha, \beta)$ plane (B). Optimal
flow with fixed enstrophy in the $(x,y)$ plane (C) and the
$(\alpha, \beta)$ plane (D). Here we use $\beta_{min} = -2$, 
$\beta_{max} = 2$.}
\label{fig:PlateEnstrophy}
\end{figure}

For the optimal flow over a heated plate in section \ref{sec:plate}, the domain is infinite. 
We have $h = \frac{\pi}{2}\left(\cosh(\pi \beta) - \cos(\pi \alpha)\right)^{-1}$.
As $(\alpha,\beta)\to (0,0)$, $h \sim \|(\alpha,\beta)\|^{-2}$. Thus 
the small-$Pe$ eigenvalue problem is singular. Solving the problem 
numerically in the $(\alpha,\beta)$ rectangle, we find that
the solutions have infinite kinetic energy and do not satisfy
$\psi \to 0$ as $(\alpha,\beta)\to (0,0)$ (or $x + iy \to \infty$). To examine the situation further, 
we consider a regularized version of
the problem, in which the domain is cut off by a large semi-circle of radius
$R$ on which $T = 0$ (see figure \ref{fig:PlateEnstrophy}A). We compute 
$\alpha$ and $\beta$ for this domain by modifying the unbounded
definition (\ref{abplate}):
\begin{align}
\alpha + i\beta =  \frac{-i}{\pi}\log{\left(\frac{z-1/2}{z+1/2}\right)}
-i\sum_{k = 1}^\infty a_k z^k.
\label{abplateMod}
\end{align}
\nn We have added a power series, which gives a general representation of
an analytic function which converges inside the disk of radius $R$.
We truncate the series at $k = N$ and choose the $a_k$ so that 
the $\sin{k\theta}$-components of $\alpha$ are zero 
on $|z| = R$. We find that
$a_k$ converge rapidly to 0 with increasing $k$, so just a 
few terms are needed to obtain $\sim 10^{-12}$ precision for $R = 10$.   

In figure \ref{fig:PlateEnstrophy} we compare the optimal flows with
fixed kinetic energy and enstrophy in the small-$Pe$ limit with $R = 10$.
The flow with fixed kinetic energy (panel A) is almost identical to that in the
unbounded case (figure \ref{fig:PlateFlowsFig}A), because the flow
is very weak far from the plate in that case. Panel B shows the same
flow in the $(\alpha,\beta)$ plane. The optimal flow with fixed 
enstrophy (panel C) has a single vortex whose size is roughly that of the
domain. Panel D shows this flow in the $(\alpha,\beta)$ plane. The vorticity is 
concentrated near $(\alpha,\beta) = (0,0)$, where $h$ is largest.

In summary, the optimal flows with fixed enstrophy have vortices which are
much larger in scale than those for flows with fixed kinetic energy.
In the examples shown here, the fixed-enstrophy flows 
are of the same size as the physical domain, while the fixed-energy
flows are of the order of the size of the hot surface. 


\section{Summary and possible extensions}\label{Conclusion}

We have shown that a certain class of optimal flows
for heat transfer can be generalized to a wide
range of geometries using a change of coordinates. We presented solutions
in three basic situations: the exterior cooling of
a cylinder, a hot flat plate embedded in a cold surface, and
a channel with hot interior and cold exterior.
Steady 2D flows were sufficient to present the basic idea, 
but unsteady flows can be addressed by retaining the
unsteady term in the advection-diffusion equation. It seems likely 
that the coordinate change would be useful in certain 3D geometries as well,
though these problems have not yet been solved.



\bigskip
We acknowledge helpful discussions with Professor Charles R. Doering.

\appendix
\section{Alternative formula for $Q$}\label{QAlt}
Here we show that the two formulas for $Q$, (\ref{Conda}) and (\ref{Q2}), are equal.
Defining
the heat flux vector as $\mathbf{q} = \mathbf{u}T - \nabla T = -T\nabla^\perp \psi - \nabla T$,
equation (\ref{T}) can be written
\begin{align}
0 = \nabla \cdot \mathbf{q} = \frac{1}{h^2}\left[\partial_\alpha(h\mathbf{q}\cdot \hat{\mathbf{e}}_\alpha) +
\partial_\beta(h\mathbf{q}\cdot \hat{\mathbf{e}}_\beta)\right].
\end{align}
\nn We integrate over $\beta$:
\begin{align}
\int_{\beta_{min}}^{\beta_{max}} \partial_\alpha(h\mathbf{q}\cdot \hat{\mathbf{e}}_\alpha) d\beta
=-\int_{\beta_{min}}^{\beta_{max}} \partial_\beta(h\mathbf{q}\cdot \hat{\mathbf{e}}_\beta) d\beta
=h\mathbf{q}\cdot \hat{\mathbf{e}}_\beta|_{\beta_{min}}^{\beta_{max}} = 0. \label{flux}
\end{align}
\nn The last equality holds for the annulus because then $\mathbf{q}$ is periodic in $\beta$. The equality
also holds for all the other problems in this work, for which the $\beta$ boundaries
are solid insulating walls. At such boundaries we have $\mathbf{q}\cdot \hat{\mathbf{e}}_\beta = 0$.
Defining
\begin{align}
f(\alpha) = -\int_{\beta_{min}}^{\beta_{max}} h\mathbf{q}\cdot \hat{\mathbf{e}}_\alpha d\beta
\end{align}
and interchanging the derivative and integral in (\ref{flux}) we have
$f'(\alpha) = 0 \Rightarrow f(\alpha)$ = constant. The constant is
\begin{align}
f(1) = -\int_{\beta_{min}}^{\beta_{max}} h\partial_n T|_{\alpha = 1} d\beta = -\int_{\alpha = 1} \partial_n T ds = Q.
\end{align}
\nn using no flow penetration ($\mathbf{u}\cdot \hat{\mathbf{e}}_\alpha = 0$) along the hot boundary. 
Because $f(\alpha)$ is constant we also have
\begin{align} 
Q = \int_0^1 f(\alpha) d\alpha = -\int_0^1 \int_{\beta_{min}}^{\beta_{max}} h\mathbf{q}\cdot \hat{\mathbf{e}}_\alpha d\beta d\alpha
= -\int_{\beta_{min}}^{\beta_{max}} \int_0^1 h\mathbf{q}\cdot \hat{\mathbf{e}}_\alpha d\alpha d\beta
\end{align}
\nn Using $\mathbf{q}\cdot \hat{\mathbf{e}}_\alpha = u_\alpha T - \frac{1}{h}\partial_\alpha T$,
\begin{align} 
Q =  \Delta \beta - \int_{\beta_{min}}^{\beta_{max}} \int_0^1 h u_\alpha T d\alpha d\beta 
=  \Delta \beta - \int_{\beta_{min}}^{\beta_{max}} \int_0^1 \partial_\beta \psi\, T \,d\alpha d\beta. 
\end{align}

\bibliographystyle{jfm}
\bibliography{OptimalChannelFlow}

\end{document}